\documentclass[useAMS,usenatbib]{mnras}
\usepackage{amsmath,psfig,epsfig,graphics}
\usepackage{graphicx}
\usepackage{epstopdf}
\usepackage{float}

\def\chandra    {{\em Chandra}\/}

\def\xmm        {XMM-{\em Newton}\/}

\def\vla        {{\em VLA}\/}

\newcommand{\RNum}[1]{\uppercase\expandafter{\romannumeral #1\relax}}

\begin{document}

\title[SHOCKING FEATURES IN 3C89]{SHOCKING FEATURES IN THE MERGING GALAXY CLUSTER RXJ0334.2-0111}

\author[Dasadia et al.]
{Sarthak Dasadia${}^1$\thanks{E-mail: sbd0002@uah.edu}, Ming Sun${}^1$\thanks{E-mail: ms0071@uah.edu}, Andrea Morandi${}^1$, Craig Sarazin${}^2$, Tracy Clarke${}^3$, \newauthor Paul Nulsen${}^4$, Francesco Massaro${}^5$, Elke Roediger${}^6$, Dan Harris${}^7$\thanks{Dan Harris passed away on December 6th, 2015. His career spanned much of the history of radio and X-ray astronomy. His passion, insight and contributions will always be remembered.}, Bill Forman${}^7$\\
\\
$^{1}$ Physics Department, University of Alabama in Huntsville, Huntsville, AL 35899, USA\\
${}^2$ Department of Astronomy, University of Virginia, Charlottesville, VA 22904, USA\\
${}^3$ Naval Research Laboratory, Washington, DC 20375, USA\\
${}^4$ ICRAR, University of Western Australia, 35 Stirling Hwy, Crawley, WA 6009, Australia\\
${}^5$ University of Turin, I-10125 Torino, Italy\\
${}^6$ E.A. Milne Centre for Astrophysics, Department of Physics \& Mathematics, University of Hull, Hull, HU6 7RX, UK\\
${}^7$ Smithsonian Astrophysical Observatory, Cambridge, MA 02138, USA
}

\pagerange{\pageref{firstpage}--\pageref{lastpage}} \pubyear{2002}

\maketitle

\label{firstpage}

\begin{abstract}

We present a 66 ksec \textit{Chandra} X-ray observation of the galaxy cluster RXJ0334.2-0111. This deep observation revealed a unique bow shock system associated with a wide angle tail (WAT) radio galaxy and several intriguing substructures. The temperature across the bow shock jumps by a factor of $\sim$ 1.5 (from 4.1 keV to 6.2 keV), and is consistent with the Mach number $M$ = 1.6$_{-0.3}^{+0.5}$. A second inner surface brightness edge is a cold front that marks the border between infalling subcluster cool core and the ICM of the main cluster. The temperature across the cold front increases from 1.3$_{-0.8}^{+0.3}$ keV to 6.2$_{-0.6}^{+0.6}$ keV. We find an overpressurized region $\sim$ 250 kpc east of the cold front that is named ``the eastern extension (EE)''. The EE may be a part of the third subcluster in the ongoing merger. We also find a tail shaped feature that originates near the bow shock and may extend up to a distance of $\sim$ 1 Mpc. This feature is also likely overpressurized. The luminous FR-I radio galaxy, 3C89, appears to be the cD galaxy of the infalling subcluster. We estimated 3C89's jet power from jet bending and the possible interaction between the X-ray gas and the radio lobes. A comparison between the shock stand-off distance and the Mach number for all known shock front/cold front combinations suggests that the core is continuously shrinking in size by stripping. 

\end{abstract}

\begin{keywords}
galaxies: clusters: individual: RXJ0334.2-0111, 3C89 -galaxies: clusters: intracluster medium -X-rays: galaxies: clusters -galaxies: jets
\end{keywords}

\section{Introduction}

\begin{table*}
\protect\caption{Properties of RXJ0334.2-0111 }
\begin{tabular}{|c|c|c|c|c|cl}
\hline 
$kT^{a}$ & R.A. & Decl. & Redshift  & $N_{\rm H}$ & $log$($L_{X, bol}$/erg sec$^{-1}$)$^b$\\(keV) & (J2000) & (J2000)& $z$ & ($\times$ 10$^{20}$ cm$^{-2}$) & \tabularnewline
\hline 
4.9 $\pm$ 0.3 & +03:34:15.6 & -01:10:56 & 0.1386 & 7.6 & 45.03\tabularnewline
\hline 
\end{tabular}
\\
$^{a}$ The spectroscopic temperature in the radial range 0.15 - 0.75 $r_{500}$.\\
$^{b}$ The bolometric luminosity within $r_{500}$. 
\end{table*}

\begin{table*}
\protect\caption{Most luminous galaxies in the cluster}
\begin{tabular}{|c|c|c|c|c|cl}
\hline 
Galaxy & R.A. & Decl. & Velocity$^{a}$ & $log$($L_{Ks}^{b}$/$L_{\bigodot}$) & Mag$^{c}$\\& (J2000) & (J2000)& (km sec$^{-1}$)& \tabularnewline
\hline 
2MASX J03341605-0111297 (BCG 1) & +03:34:16.1 & -01:11:29 & 41620 $\pm$ 10 & 11.99 & 12.82\tabularnewline
\hline 
3C89 (BCG 2) & +03:34:15.6 & -01:10:56 & 41913 $\pm$ 6 & 11.94 & 13.35\tabularnewline
\hline 
2MASX J03343406-0109527 (BCG 3) & +03:34:34.1 & -01:09:52 & 41650 $\pm$ 10 & 11.94 & 13.20\tabularnewline
\hline 
\end{tabular}
\\
$^{a}$ {\em SDSS} velocity. The velocity difference between BCG 1 and 2 is $\sim$ 300 km sec$^{-1}$ [$\sim$ 70 km sec$^{-1}$ if the velocity of the BCG 2 is adopted from \cite{s3}]. \\
$^{b}$ {\em 2MASS} Ks band luminosity. \\
$^{c}$ {\em WISE} 3.4 $\mu$m magnitude 
\end{table*}

Galaxy clusters are the most massive gravitationally collapsed objects in the universe. In the hierarchical scenarios of the large-scale structure formation of the universe, they form by subcluster mergers and infall. Major mergers inject tremendous amounts of energy ($\sim$ 10$^{64}$ ergs) into the intra-cluster medium (ICM), triggering shocks, generating turbulence and accelerating particles \citep[e.g.,][]{m2}. Both observations and simulations show that cluster mergers change the physical characteristics of the ICM. This provides an excellent opportunity to understand some important aspects of the ICM physics, such as thermal conduction, viscosity, self-interaction of dark matter and particle acceleration \citep[e.g.,][]{s2,m3,m2,r1,b1,z1}.

\begin{figure*}
 \centering
 \hbox{\hspace{-5px}\includegraphics[scale=.95]{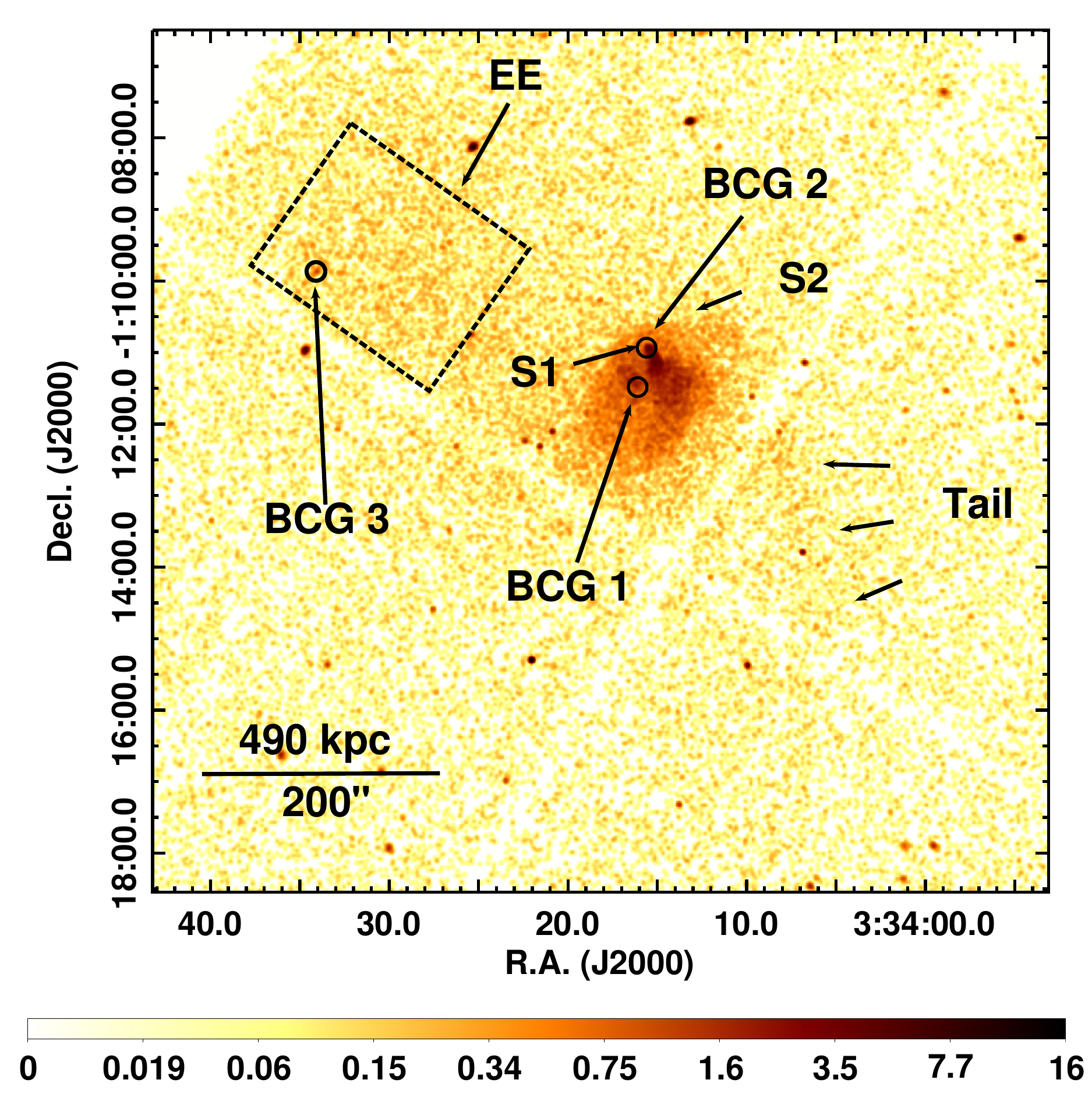}}
 \caption{The count image of the merging galaxy cluster RXJ0334.2-0111 in the 0.7 - 7 keV band, binned by 1" and smoothed with a Gaussian 3-pixel kernel. The image shows two sharp edges (S1 \& S2) in surface brightness with additional substructures associated with the merger. The locations of the three most luminous cluster galaxies in the near-infrared (NIR) are marked (see Table 2). We label substructures discussed in this paper.}
 \end{figure*}

Dynamical activities triggered by mergers often produce contact discontinuities between gas of different entropies that can be seen as surface brightness edges in X-ray observations. After the launch of \chandra, with its sub-arcsec resolution, many previously unseen hydrodynamical phenomena have been revealed, such as cold fronts and bow shocks caused by mergers. Cold fronts in mergers are ubiquitous and can be produced in primarily two ways. During a merger, a front can appear at the border separating the infalling cool core of the merging subcluster from the hot cluster atmosphere. In such a region, gas on the denser, downstream side is colder than that on the upstream side. In a minor merger, the dense cool core oscillates in the gravitational potential well forming a sloshing cold front. Cold fronts provide a unique laboratory to study transport processes e.g., thermal conduction and viscosity. When a dense core moves through a more rarified intracluster medium, the gas flow may drive Rayleigh-Taylor or Kelvin-Helmholtz instabilities.  Observations of the growth rates of these instabilities can be used to estimate the cluster-scale magnetic
field and the transport properties of the ambient gas \citep[e.g.,][]{m2}.

Unlike cold fronts, bow shocks in merging clusters are relatively rare, e.g., the ``Bullet Cluster'' 1E 0657-56 \citep{m4}, Abell 520 \citep{m3}, Abell 2146 \citep{r3}, Abell 754 \citep{m1}, Abell 2744 \citep{o3}, RX J0751.3+5012 \citep{r2}, Abell 521 \citep{b5} and Abell 2034 \citep{o2}. Observationally, the bow shocks are visible as edges in X-ray surface brightness. The projection effects and other irregularities in the image, such as the presence of substructures, bring observational challenges to detect bow shocks. Thus, for a prominent shock detection, the compression of the gas has to be near the plane of sky \citep[e.g.,][]{o2}. Additionally, the gas velocities in the merger are of the order of 10$^{3}$ km sec$^{-1}$ which implies a shock of Mach number $M \leq$ 3. This means such a shock would not form a huge temperature contrast and would require accurate spectroscopy to estimate the gas velocity on the plane of sky \citep[e.g.,][]{m5, m2}. Thus, deep X-ray observations are required for the analysis of faint merger features.

During our systematic search of merger shocks in the \chandra\ archive, the RX J0334.2-0111 cluster was selected for a sharp edge ahead of the moving cool core of 3C89. The observation for the original selection (ObsID: 12724) revealed comparable temperature difference across this edge indicating the presence of a discontinuity caused by the merger. To further investigate, we obtained longer \chandra\ observations in 2012. In this paper, we present the results from our new 66 ksec \chandra\ data of the merging galaxy cluster RX J0334.2-0111.

RX J0334.2-0111 was initially detected by the ROSAT all-sky survey at the redshift 0.1386. It was observed by \chandra\ due to the radio galaxy 3C89 in the cluster, as part of the 3C galaxy snapshot program \citep{m13,m14}. The properties of the RX J0334.2-011 are listed in Table 1. The cluster hosts a luminous FR-I radio galaxy, 3C89, which has a wide angle tail (WAT) like morphology.  The properties of the three most NIR luminous galaxies in NIR in the cluster are listed in Table 2 and their locations are shown in Fig. 1.  Compared to 3C89 (BCG 2), 2MASX J03341605-0111297 (BCG 1) is closer to the geometric center of the cluster and the most luminous in the NIR. Although 3C89 is not the most NIR luminous galaxy in the cluster, it is likely the dominant galaxy of the infalling subcluster. The velocity difference between two galaxies is only \ensuremath{\sim} 300 km sec$^{-1}$ from {\em SDSS}. \cite{s3} estimated the velocity of 3C89 to be $\sim$ 41551 km sec$^{-1}$, which reduces the velocity difference between two galaxies to $\sim$ 70 km sec$^{-1}$. This implies that the merger is taking place near the plane of sky. The 2MASX J033434406-0109527 (BCG 3)  is located $\sim$ 700 kpc east from 3C89. With luminosity comparable to that of a typical brightest cluster galaxy (BCG), BCG 3 is likely to be the BCG of the third subcluster.

The paper is structured as follows: the \textit{Chandra} data reduction and background modeling are presented in section 2.  Spatial features of the merger and radial profiles are in section 3. In section 4, the spectral properties and the projected temperature map are discussed. 3C89's AGN and jet properties are discussed in section 5. The Chandra data also reveal other intriguing features related
to the merger, which are discussed in section 6. In sections 7 and 8, we discuss the new results and present our conclusions. Throughout this paper, we assume $H_{0}$= 70 km sec$^{-1}$ Mpc$^{-1}$, $\Omega_{M}$ = 0.3, $\Omega_{\Lambda}$ = 0.7. For this cosmology, at z = 0.1386, an angular size of 1 arcsec corresponds to a distance of 2.45 kpc. All error bars reported show 1$\sigma$ confidence interval unless otherwise noted.

\section{Chandra Data Reduction}

\begin{table*}
\protect\caption{Summary of the \chandra\ observations (PI: M. Sun)}
\begin{tabular}{|c|c|c|c|c|}
\hline 
ObsID & R.A. & Dec. & Total Exposure & Effective Exposure\\& (J2000)& (J2000)& (ksec)& (ksec)\tabularnewline
\hline 
14028 & +03:34:16.00 & -01:11:17.40 & 38.21 & 37.71\tabularnewline
\hline 
14378 & +03:34:16.00 & -01:11:17.40 & 28.97 & 28.59\tabularnewline
\hline 
\end{tabular}
\end{table*}

RX J0334.2-0111 was observed by \textit{Chandra} with the Advanced CCD Imaging Spectrometer (ACIS) in the Very Faint (VFAINT) mode for a total exposure of 67 ksec. The details of the \chandra\ observations are summarized in Table 3. All observations were taken with the ACIS-I. For data analysis, we used Chandra Interactive Analysis of Observations (CIAO) v4.6 and calibration database (CALDB) v4.6.1.1 from the \textit{Chandra} X-ray Center\footnote{Our results are not affected by the update in the CALDB v4.6.8.}. For each observation, a new level = 2 reprocessed event file was made using {\tt CHANDRA\_REPRO} script with VFAINT mode correction. This process accounts for afterglows, charge transfer inefficiency, bad pixels and gain correction.  {\tt CHECK\_VF\_PHA} was set to ``yes'' to remove bad events that are likely associated with cosmic rays which are more easily detected in VFAINT mode. A light curve of the source-free region was then analyzed to identify any background flares or other bad events. The {\tt DEFLARE} script was used to filter good time intervals. The observations were not affected by background flares. 

A point spread function map was extracted at 1.5 keV enclosing 40\% of total counts to detect point sources. Approximately 100 point sources were detected using {\tt WAVDETECT} and removed from the analysis. An exposure map was then made to account for quantum efficiency (QE), vignetting and effective area at 0.7 - 2.0 keV. A weighted exposure map was created to account for the energy dependence of the effective area using the absorbed APEC model with $N_{\rm H}$ = 7.6 $\times$ 10$^{20}$cm$^{-2}$ and $z$ = 0.1386. We have also examined the X-ray absorption column density towards the cluster. The H\RNum{1} observations by \cite{k1} give a column density of 7.0 $\times$ 10$^{20}$cm$^{-2}$. The relation between the total hydrogen column density, the atomic hydrogen column density and dust extinction, derived by \cite{w1}, predicts $N_{\rm H}$ = 9.75 $\times$ 10$^{20}$cm$^{-2}$. We fitted the spectrum extracted from a circular region centered on the X-ray peak with a radius of 5 arcmin. The absorption parameter was kept free to obtain the best-fit value of $N_{\rm H}$ = 7.6 $\times$ 10$^{20}$cm$^{-2}$. We report around 8 - 10\% systematic error in the best fit parameters like temperature, abundance and normalization that might have occurred by changing the $N_{\rm H}$ value.

The standard stowed background file for each chip in each observation was reprojected to match the time dependent aspect solution for the observation. The stowed background files used for the analysis are: acis?D2009-09-21bgstow\_ctiN0001.fits, \{? = 0, 1, 2, 3, 6\}. The stowed background was normalized to match the hard X-ray background count rate in the 9.5 - 12.0 keV energy band. The correction needs to be applied to the particle component of the total background that is dominant at high energies and has a significant impact on gas temperature fitting. This correction was around 10\% for each data set.

Proper background modeling is important for spectral analysis especially in regions of low surface brightness. The hard cosmic X-ray background (CXB) was modeled by an absorbed powerlaw with a photon index of 1.5. The soft CXB component is described by two thermal components at solar abundance and zero redshift, one absorbed component fixed at temperature 0.25 keV and another unabsorbed at fixed temperature 0.1 keV \citep[e.g.,][]{s1, s4}. To test the background model, we compared the derived soft CXB flux with the RASS R45 values around the cluster \citep[see the detailed discussion in][]{s4}. The R45 count rate in the 0.47 - 1.21 keV energy band around this cluster is (136.7 $\pm$ 7.3) $\times$ 10$^{-6}$ counts sec$^{-1}$ arcmin$^{-2}$, which predicts a soft CXB flux of $\sim$ 5 $\times$ 10$^{-12}$ ergs sec$^{-1}$ cm$^{-2}$ deg$^{-2}$. We find an observed soft CXB flux of 5.0 $\times$ 10$^{-12}$ ergs sec$^{-1}$ cm$^{-2}$ deg$^{-2}$, consistent with predicted value by RASS R45. The hard CXB flux in the region is 8.9 $\times$ 10$^{-12}$ ergs sec$^{-1}$ cm$^{-2}$ deg$^{-2}$, which is also consistent with the depth of our observations \citep[see details in][]{s4}. For spectral fitting we used XSPEC version 12.8 \citep[e.g.,][]{a1}. This work used AtomDB 2.0.2. Throughout the analysis we assumed the solar abundance table by \cite{a2}.

\section{Spatial Analysis}
\subsection{Image Analysis}

\begin{figure*}
\begin{center}
\hbox{\hspace{5px}
\psfig{figure=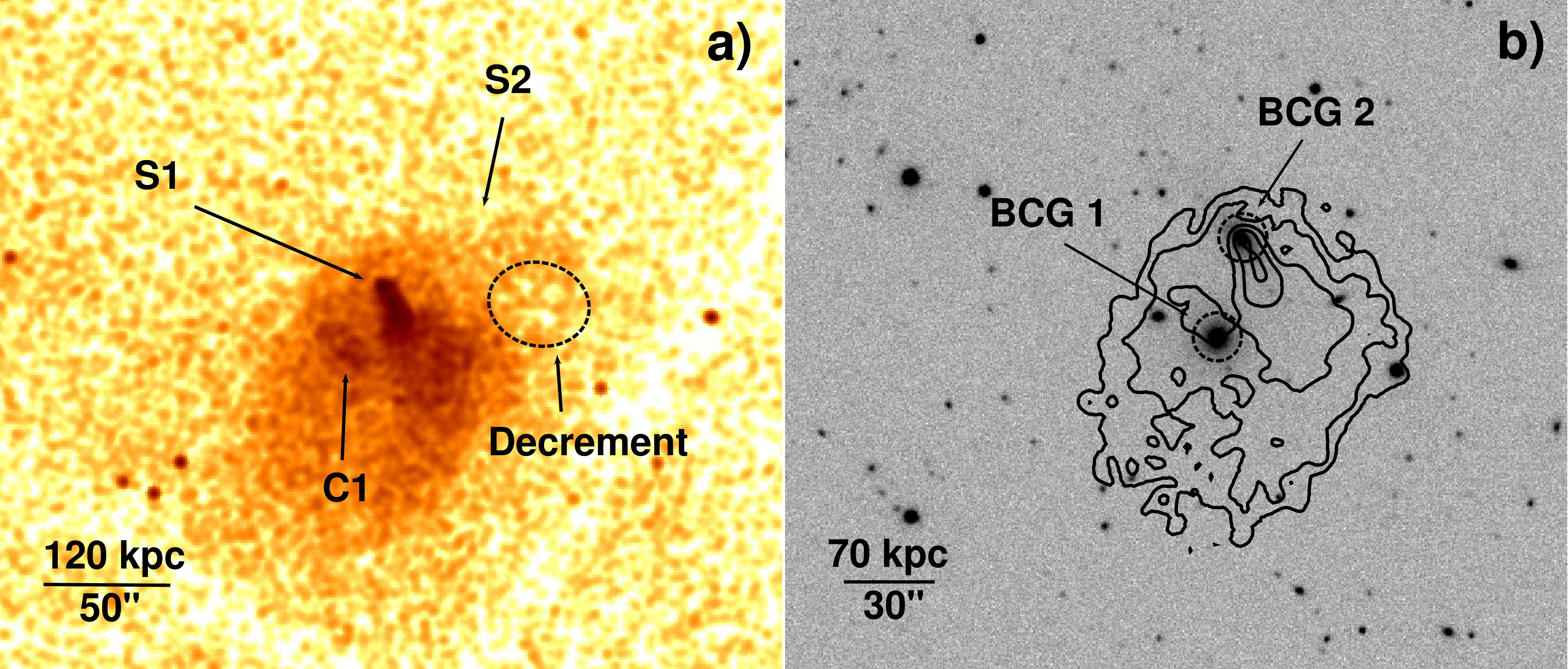,scale=0.262}
}
\vspace{10px}
\hbox{\hspace{5px}
\psfig{figure=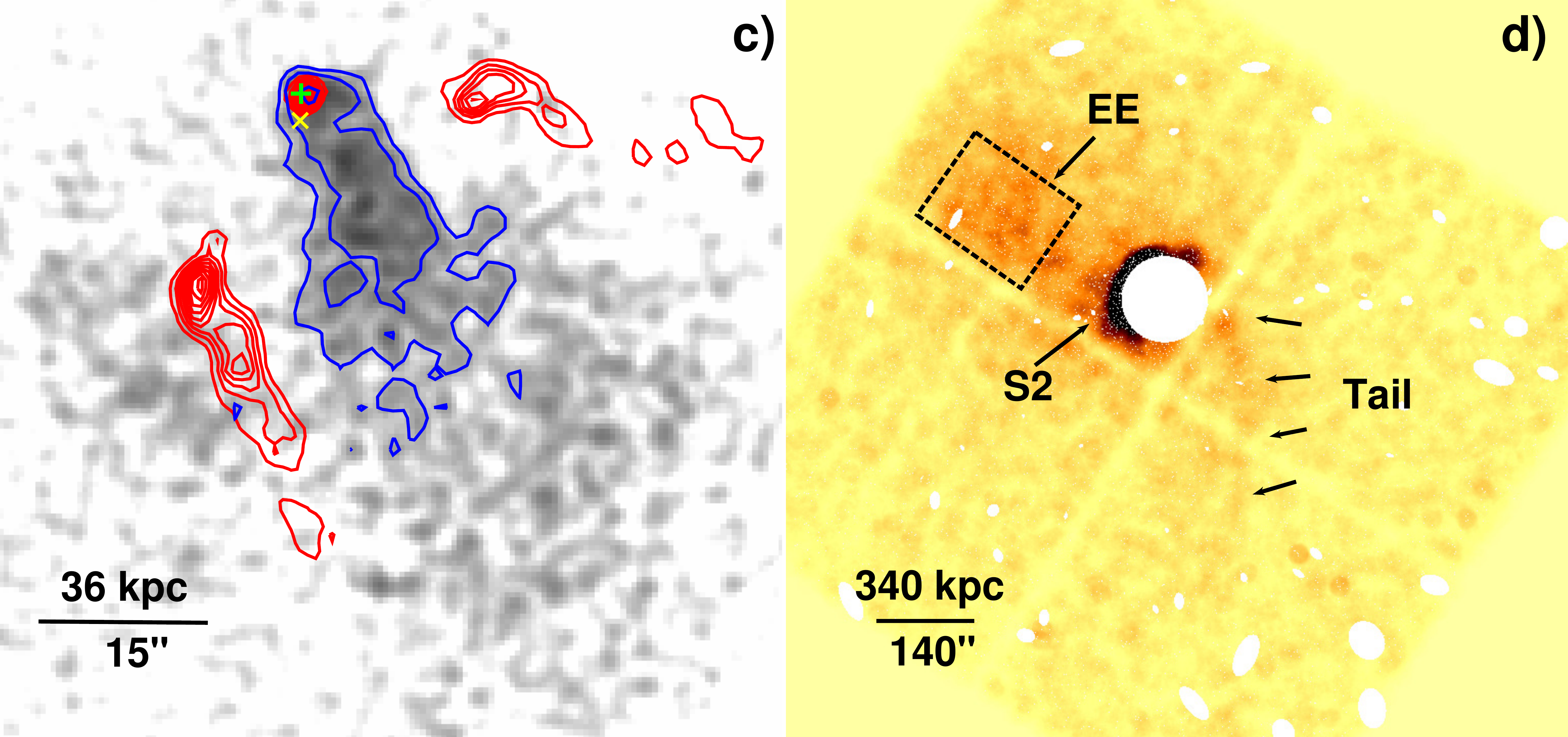,scale=0.324}
}
\caption{a) The combined exposure-corrected image of the \textit{Chandra} observations in the 0.7-2.0 keV energy band smoothed with a two dimensional Gaussian 5-pixel kernel (1 pixel = 0.492") . The arrows show locations of two surface brightness edges S1 and S2, the cool core remnant of the main cluster C1, and the Decrement. b) {\em SDSS} $i$-band image of RXJ0334.2-0111 with X-ray contours overlaid and two BCGs marked. c) The exposure-corrected \textit{Chandra} image showing a closer view of the cold front with X-ray (blue) and 1.5 GHz \vla\ radio (red) contours overlaid. The defined cluster center and the center of the surface brightness are marked by green ``+'' and yellow ``X'', respectively. d) Unsharp masked image, created by subtracting from each pixel the mean value within a radius of 30 pixel. The faint merger substructures such as EE, Tail and the S2 are indicated with arrows. The color scale runs from black to red to yellow to white, going from negative to positive deviations. The point sources and a circular region including bright cluster body was excluded. }
\end{center}
\end{figure*}

The \textit{Chandra} observations of RXJ0334.2-0111 reveal a merger between at least two subclusters. Fig. 1 shows the 0.7 - 7.0 keV count image combining both observations. The cluster gas is extended east of the merger axis, south-west (SW) to north-east (NE). The image shows the bright, cool core of the merging subcluster is being stripped of its material by ram pressure in the merger. The X-ray morphology suggests an ongoing merger, where the infalling subcluster has just passed near the center of the main cluster from the SW. We notice a sharp drop in the surface brightness around the infalling subcluster cool core (S1 in Fig. 2a). The tail following the S1 is visible over the distance of $\sim$ 73 kpc. The main cluster may have had a cool core in the past, which was disturbed by the current merger. The cool core remnant of the main cluster (C1 in Fig. 2a) can be seen as a peak in the surface brightness SE of the inner edge. There is a second surface brightness edge (S2 in Fig. 2a)  $\sim$ 54 kpc ahead of the bright inner one. The outer edge is visible over $\sim$ 590 kpc in length enclosing the cores of the both merging components. Additional features are also observed on both sides of the merger axis. There is a density enhancement $\sim$ 250 kpc east of the main cluster center, which we call ``the Eastern Extension (EE)''. There is also a ``tail'' like feature seen west of the cold front which likely extends to $\sim$ 1 Mpc from 3C89. Another interesting feature is a region with relatively low surface brightness behind the shock front on the west side of S1, which is called ``Decrement'' (Fig. 2). More detail about this feature is discussed in section 7.1.

Fig. 2a shows the smoothed \textit{Chandra} image around the cluster center. The inner and outer surface brightness edges and the main cluster core are marked by S1, S2, and C1 respectively. The image reveals a stream of gas following the dense core in the cold front. This can occur due to the stripping of outer layer gas by ram pressure generated when the cool core is moving through the hotter surroundings. Similar structures have been observed associated with dense cool cores in mergers \citep[e.g.,][]{m9,v1,s02,r5,m2,m6,k3}. Fig. 2b shows the {\em SDSS} $i$-band image with X-ray contours overlaid and both BCGs are marked with dashed circles. The two cluster cores are separated by a projected distance of $\sim$ 93 kpc. The galaxy in the north is the radio galaxy 3C89 (marked by BCG 2). The \chandra\ data reveal a weak X-ray AGN in the 2 - 8 keV band (section 5). The overlaid X-ray contours indicate a bright tail behind 3C89. This leads to an interpretation that 3C89 is likely the dominant galaxy in the merging subcluster. The second peak in the X-ray contours is consistent with the brighter elliptical galaxy 2MASX J03341605-0111297 (marked by BCG 1) which is the cD galaxy of the main cluster.  The velocity difference of these two galaxies is only $\sim{300}$ km sec$^{-1}$, which suggests that their relative motion is near the plane of sky. Fig. 2c zooms in on the cold front and the ram pressure stripped tail. The radio image from \vla\ 1.5-GHz archival data shows the AGN and the radio lobes of 3C89. The tail and the EE are clearly visible in the unsharp-masked image (Fig. 2d).  

 \begin{figure}
\centering
\hbox{\hspace{-10px}
\psfig{figure=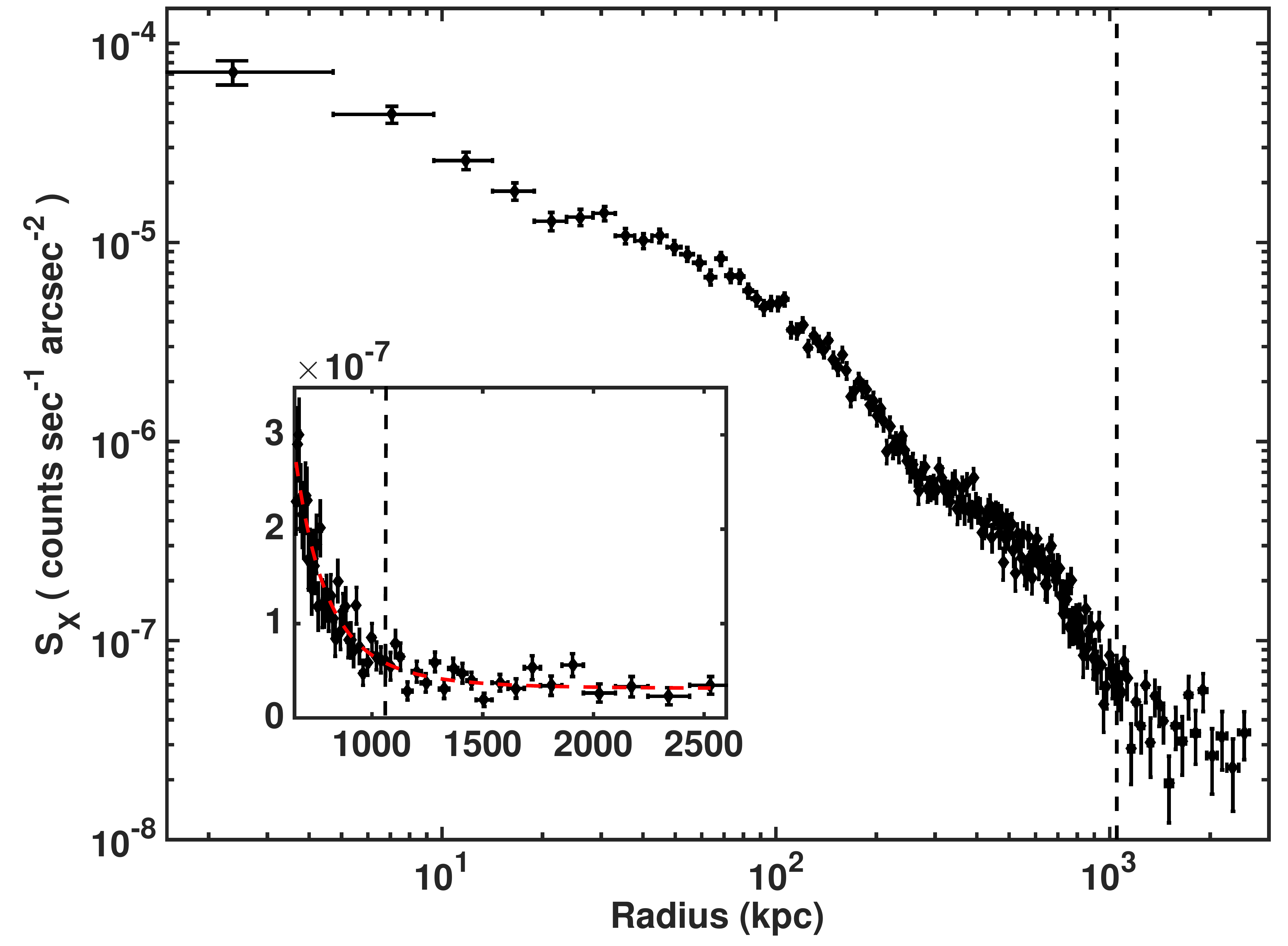,width=0.49\textwidth}
}
\caption{The azimuthally averaged surface brightness profile of the cluster in the 0.7 - 2.0 keV energy band. The $r_{500}$ for the cluster was estimated using the system temperature of 4.9 keV, and  is marked by a dashed line. The discontinuities in the surface brightness slope imply sudden changes in the density. The inset panel shows the profile for the range 500 - 2500 kpc on a linear scale. We fit the surface brightness in this region to a powerlaw plus a constant. The best-fit powerlaw index is 4.5 $\pm$ 0.4. The dashed line in red shows the best-fit model. The panel shows that the cluster emission can be traced to 3$\sigma$ up to $\sim$1 Mpc.}
\end{figure}

\subsection{Radial Profiles}

We determined the center of the cluster emission by finding the point where the derivatives of the surface brightness variation along two orthogonal directions become zero \citep{m11}. Since the derived center is only 5 kpc south of 3C89, the cluster center was taken to coincide with the nucleus of 3C89. The azimuthally averaged projected radial profile of the cluster is shown in Fig. 3. It shows that the cluster emission can be traced out to $\sim$ 1 Mpc at 3$\sigma$. The profile shows several discontinuities in the surface brightness slope. For example, the drop at $\sim$ 20 kpc shows the inner edge (S1) and the outer edge (S2) is seen at $\sim$ 70 kpc. The hump in the surface brightness profile from 250 - 700 kpc is due to the X-ray bright region present in the east (EE). The radial profile flattens beyond $\sim$ 1.5 Mpc, which is the region used to constrain the local background.

The 0.7 -- 2 keV surface brightness profiles were also determined across the edges described in section 3.1. The key regions used to investigate these features in detail are identified in Fig. 4 (the upper left panel). The regions were positioned where the surface brightness edges formed a better contrast with the surroundings. The particle background was subtracted with the normalized stowed background, while the CXB was removed in modeling. The radial bins were $\sim$ 3.5 kpc in width for the bright regions and increased to $\sim$ 21 kpc for the cluster outskirts to ensure a minimum of 20 source counts per bin.

\begin{figure*}
\begin{center}

\hbox{\hspace{-5px}
\psfig{figure=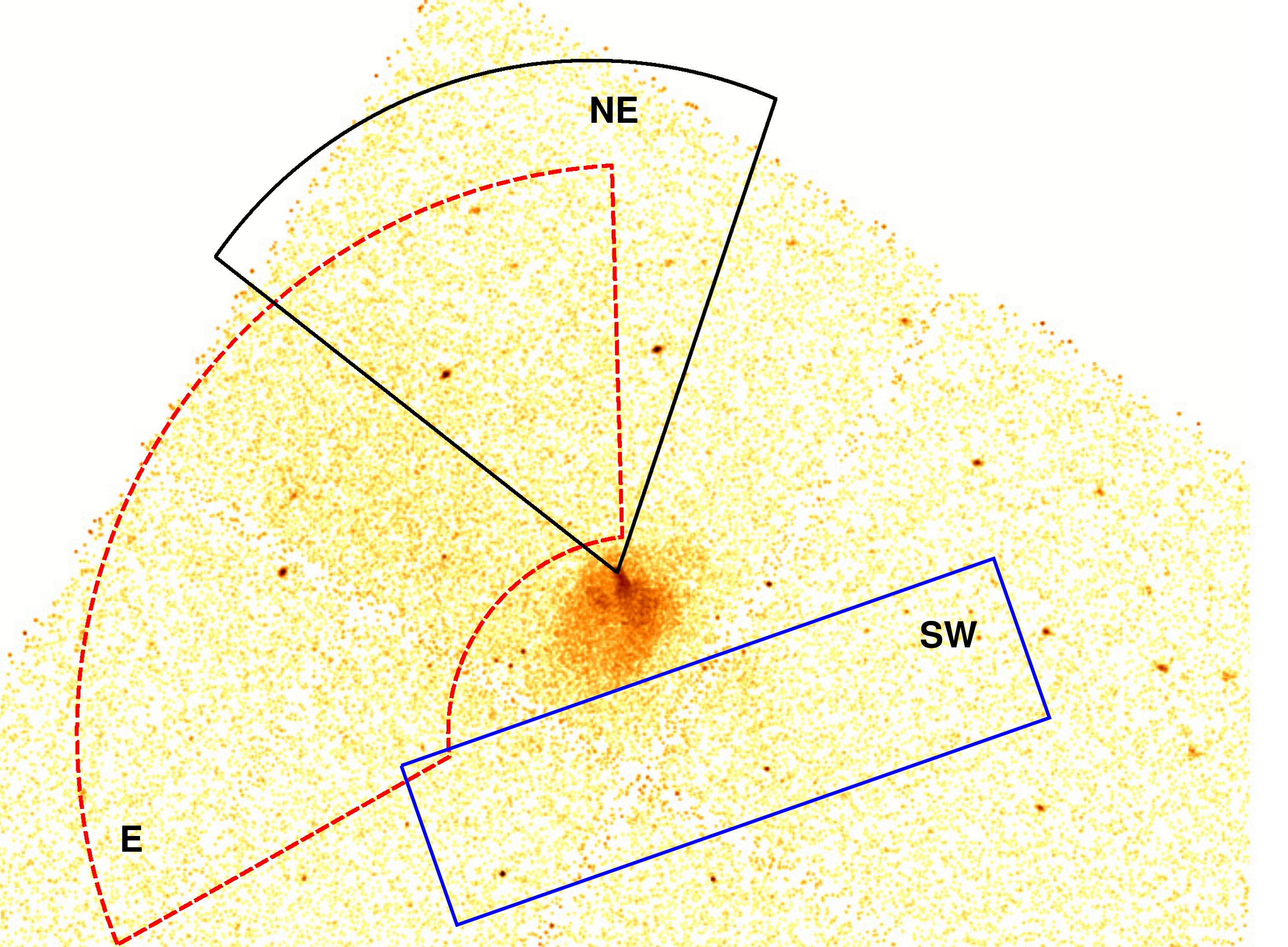,width=0.495\textwidth}
\psfig{figure=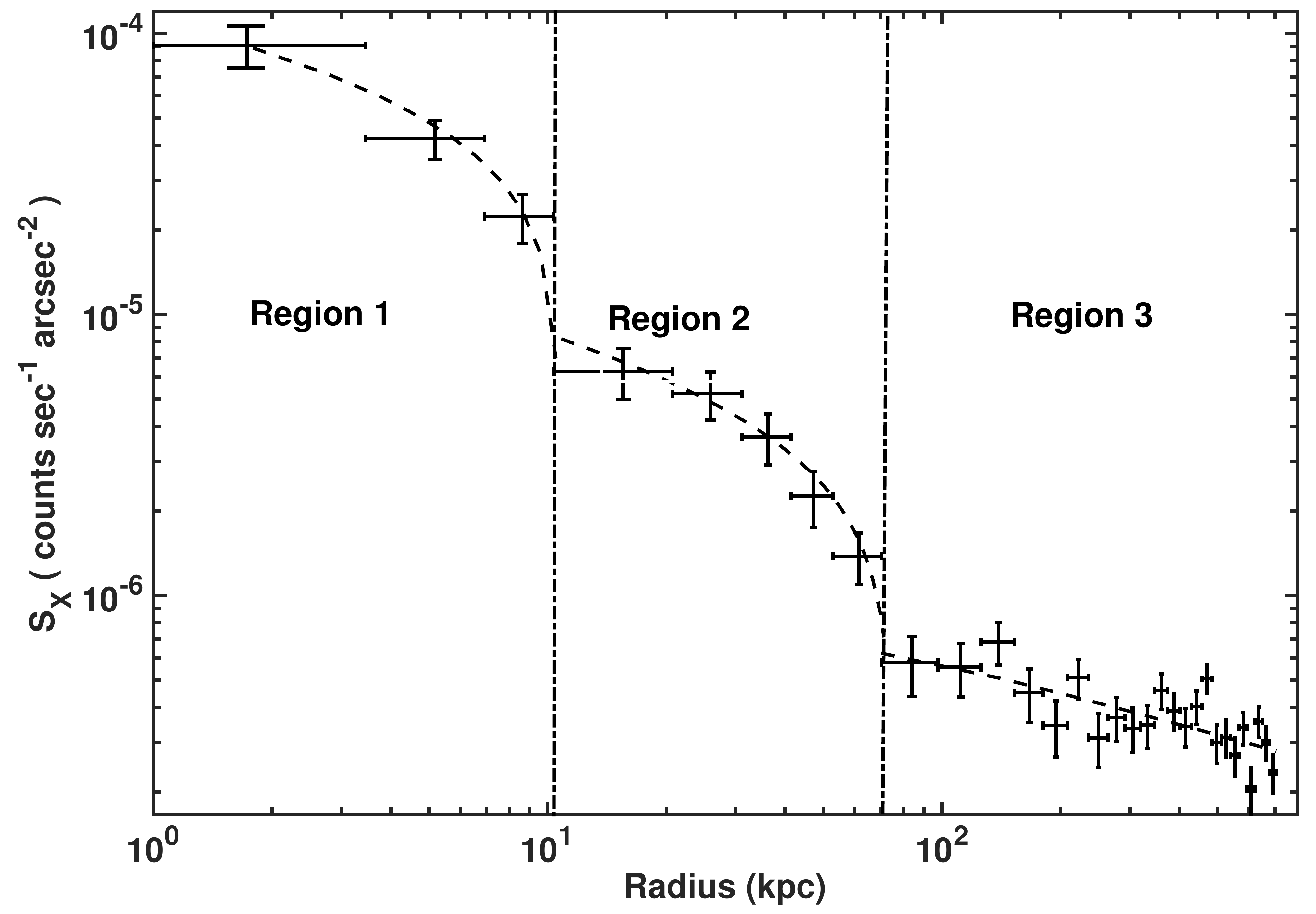,width=0.5\textwidth}
}
\vspace{10px}
\hbox{\hspace{-5px}
\psfig{figure=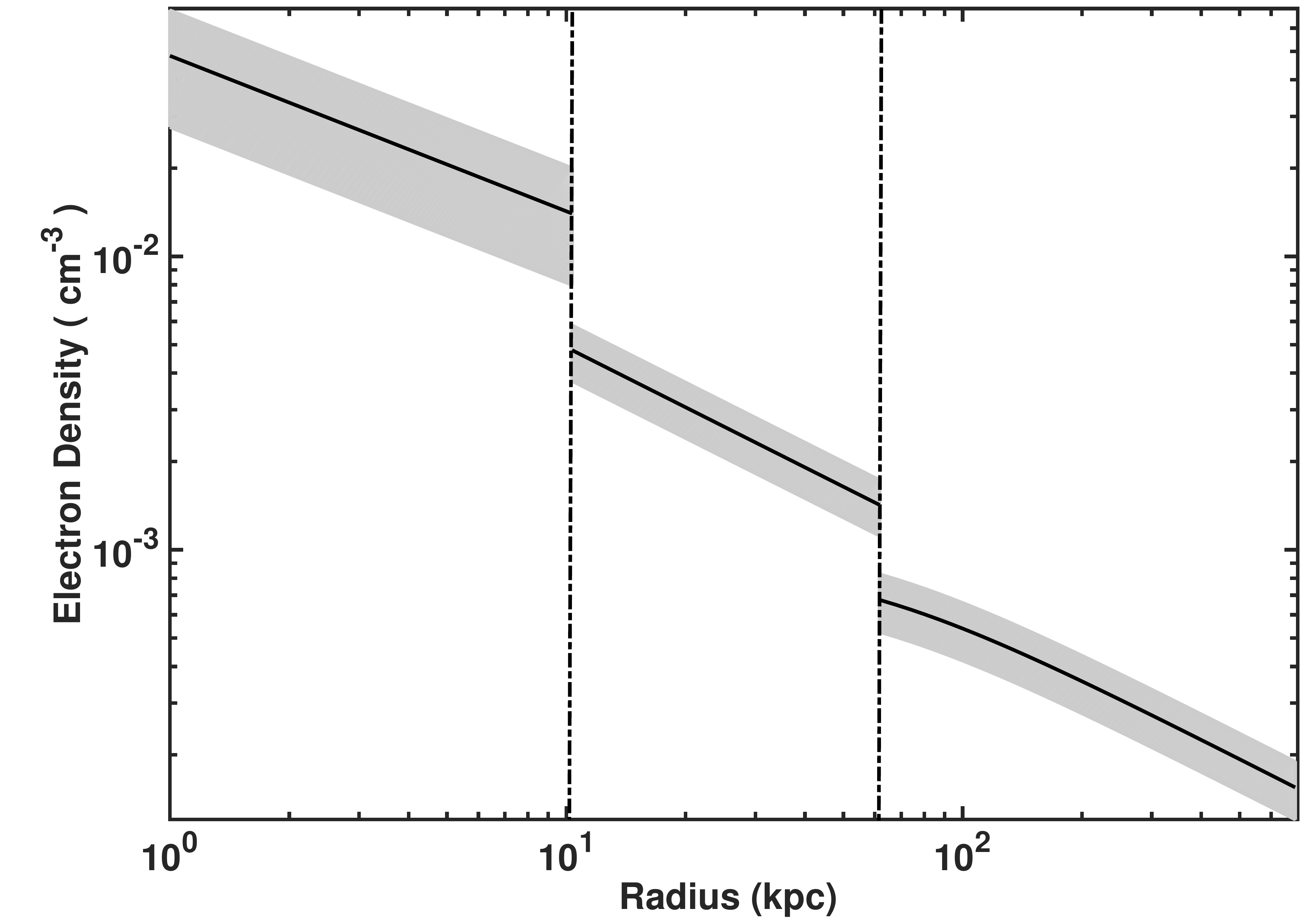,width=0.5\textwidth}
\psfig{figure=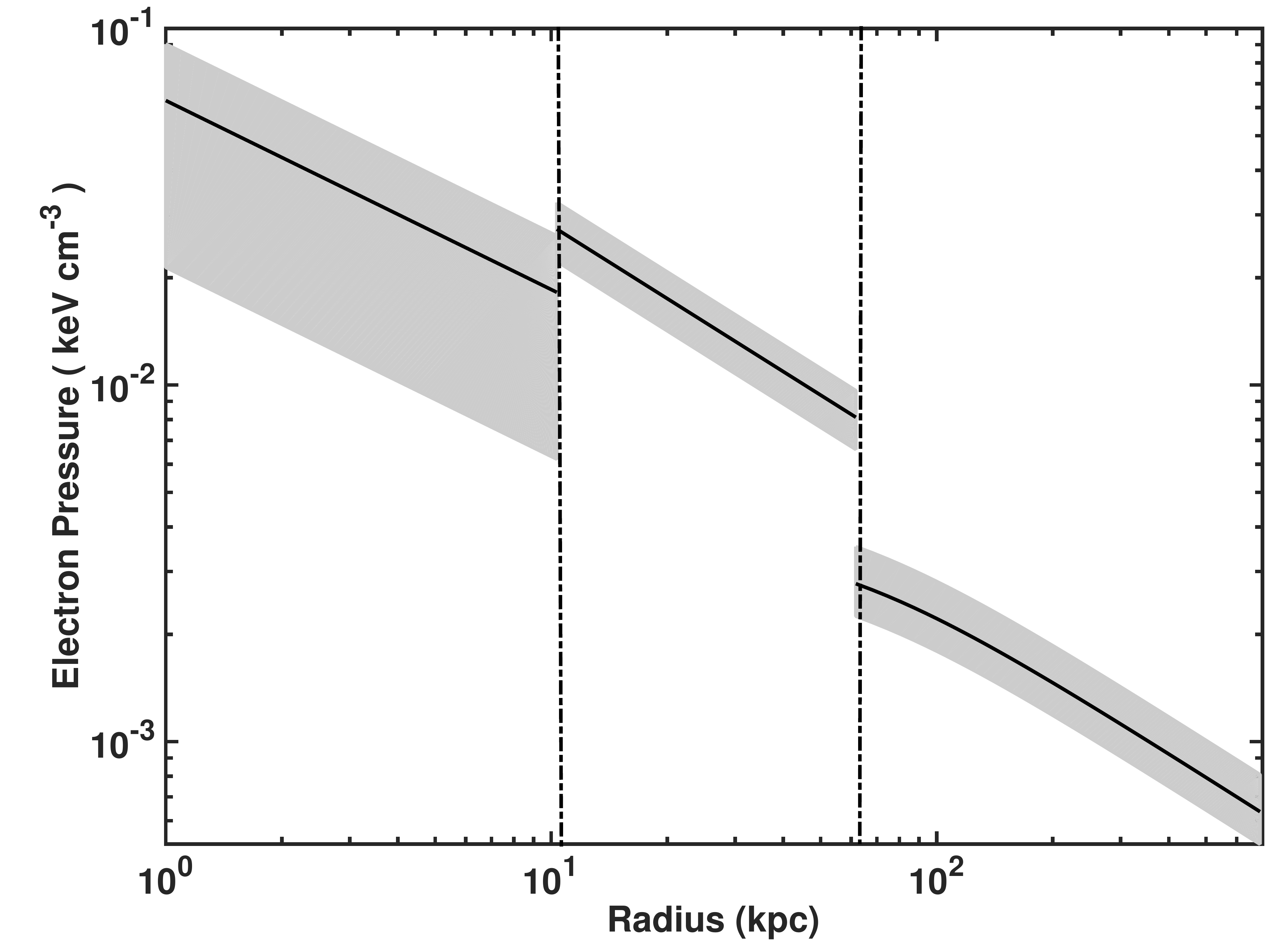,width=0.49\textwidth}
}
\vspace{10px}
\hbox{\hspace{-10px}
\psfig{figure=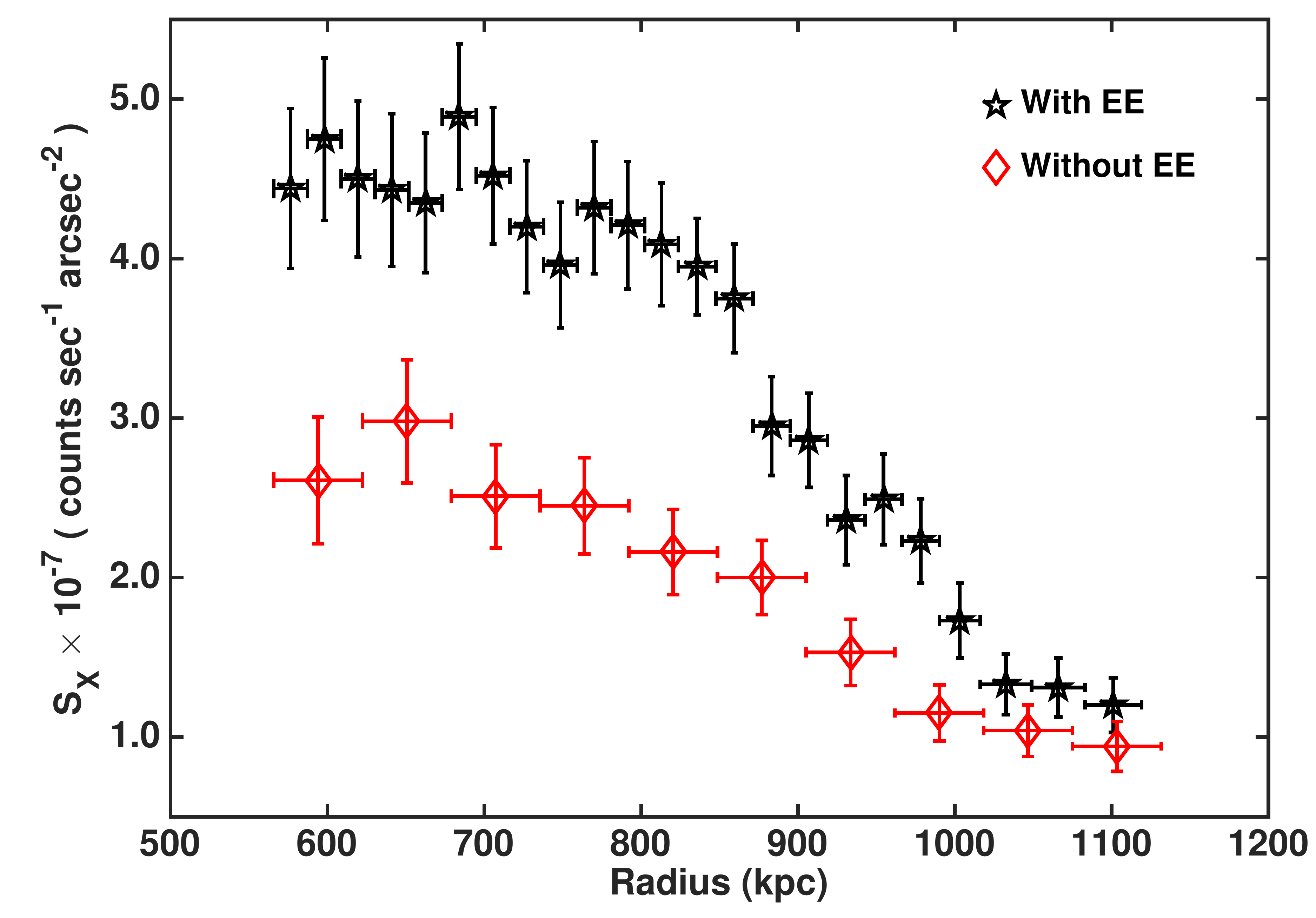,width=0.52\textwidth}
\psfig{figure=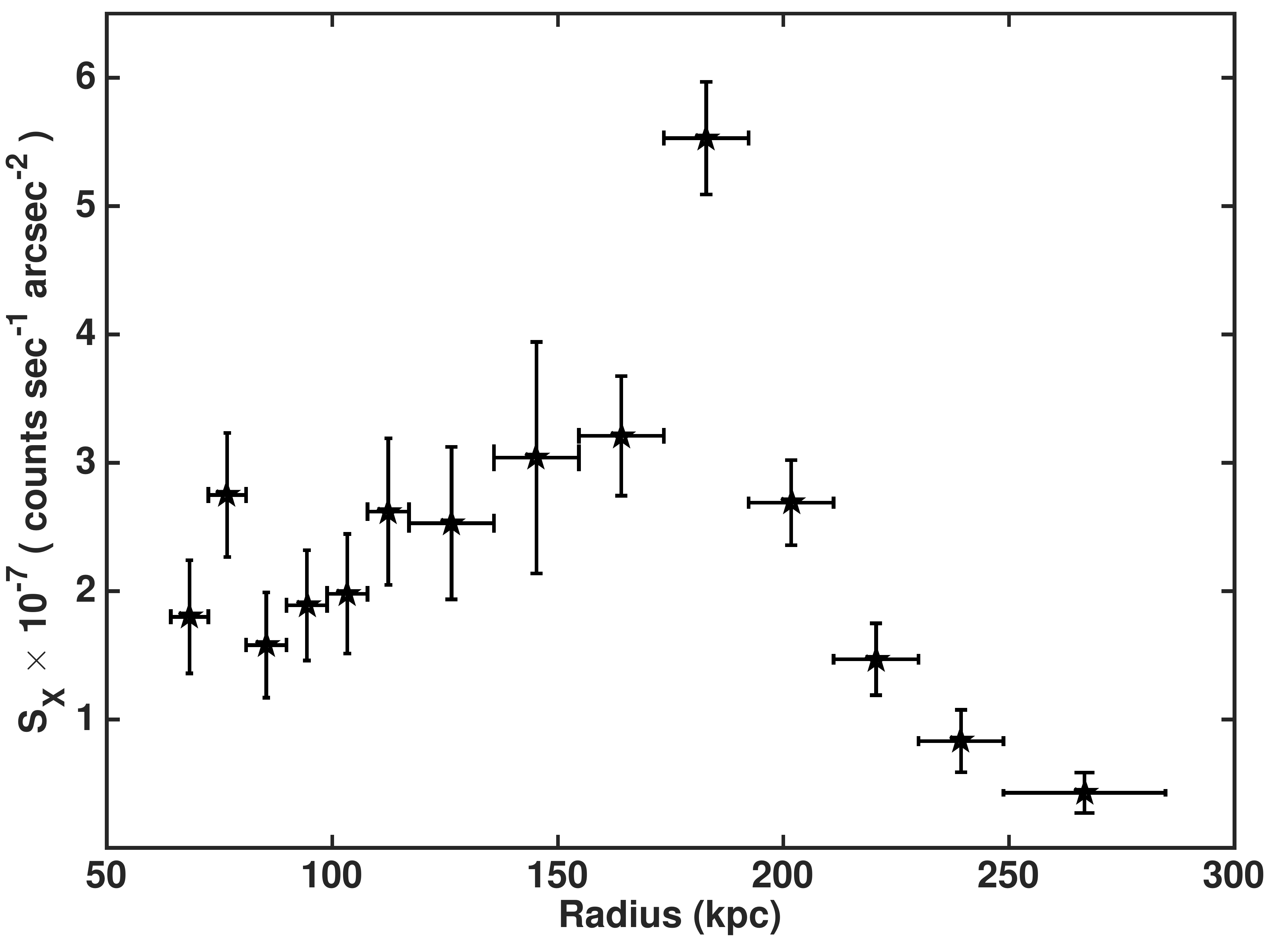,width=0.492\textwidth}
}

\caption{Upper left panel: Regions used to produce radial profiles. Upper right panel: The 0.7 - 2.0 keV radial surface brightness profile for the NE region. The three regions marked are: inside cold front (region 1), inside shock (region 2) and pre-shock region (region 3). The two vertical dash-dotted lines show the location of the cold front (left) and the shock front (right). The dashed line shows the surface brightness profile from the best-fit density model. Middle left panel: The electron density profile (with 1 $\sigma$ uncertainty) across the NE region. Middle right panel: The electron pressure profile (with 1 $\sigma$ uncertainty) across the NE region. Lower left panel: The surface brightness profile from the region E. The data points in black (stars) and red (diamonds) show the surface brightness distribution of the same radial range with and without the EE. Lower right panel: The surface brightness profile from the SW region along the long axis of the rectangle.  The short axis was taken to be roughly parallel to the Tail feature. The profile show a rise in surface brightness at the tail by $\sim 4\sigma$ above the surroundings. The deprojection of the surface brightness suggests the density rise by a factor of $\sim$ 2 relative to the surroundings. The tails appears to be $\sim$ 30 kpc wide and visible over a distance of $\sim$ 500 kpc from 3C89.}
\end{center}
\end{figure*}

The surface brightness profile for the NE region is shown in the upper right panel of Fig. 4. There are three distinct regions: the infalling merging subcluster cool core bordered by a cold front (Region 1), a shock heated region bordered by a bow shock (Region 2) and a pre-shock region (Region 3). The profile clearly shows the inner edge (S1) at 10 kpc where the surface brightness drops by a factor of $\sim$ 5 within $\sim$ 7 kpc. The outer surface brightness edge identified in Fig. 2a (S2) is detected by a drop of factor $\sim$ 2 at a radius of 60 kpc. The location of this edge ahead of subcluster core suggests it should be a bow shock. In section 4, we extract the temperature profile across these edges to determine the nature of these fronts.

The gas density distribution in region 1 \& 2 were derived by fitting corresponding surface brightness profiles to a model, where the X-ray emissivity ($\epsilon$) and radius are related by a powerlaw, $\epsilon$ $\propto$ $r^{-p}$ within each region assuming an ellipsoidal geometry as discussed in the Appendix A of \cite{k2}. The model for a region with two edges was fitted. The best-fit powerlaw index in region 1 \& 2 are 0.53 and 0.68 respectively. Using the best-fit powerlaw index $p$, we reconstructed the intrinsic emissivity distribution to obtain related density distribution, $n_e(r)$ = $[\epsilon(r)/\Lambda(T_e, Z)]^{1/2}$, where $\Lambda(T_e, Z)$ is the X-ray emissivity function which depends on electron temperature T$_e$ and abundance $Z$. The surface brightness in region 3 was fitted to a 1D $\beta$-model, $I(r)$ = $I_{0} (1+ r^{2}/r_{c}^{2})^{-3 \beta + 0.5}$. The $\beta$-model provides a good fit to the data; the best fit parameters in region 3 are $\beta$ = 0.23 and $r_c$ = 55 kpc. The best-fit $\beta$ is relatively low, which is due to the limited radial range. We used the best-fit parameters to derived the electron density model $n_e(r)$ = $n_{e0} (1+ r^{2}/r_{c}^{2})^{-3 \beta/2}$.

Fig. 4 (the middle left panel) shows the best-fit density model in all three regions. The best-fit density jump across the bow shock is $\rho_{2}$/$\rho_{1}$ = 2.2 $\pm$ 0.3, where we use suffix 1 and 2 to denote quantities upstream and downstream of the shock. We calculate the Mach number of the bow shock by applying Rankine-Hugoniot jump conditions \citep{l1}. The Mach number of the bow shock and the corresponding density jump are related by, 

\begin{equation}
M_{\rho}=\bigg[\frac{2\frac{\rho_2}{\rho_1}}{\gamma+1-(\gamma -1)\frac{\rho_2}{\rho_1}}\bigg]^{1/2},
\end{equation}

\noindent where $\gamma$ = 5/3 is the adiabatic index for monoatomic gas. \citep[e.g.,][]{m2}. The Mach number for the bow shock is $M$ = 1.9 $\pm$ 0.4. The model used to fit surface brightness profile approximates the gas distribution at edges to be spherical or elliptical, and this assumption may cause an additional systematic uncertainty in the measurement.

\section{Spectral Analysis}

\begin{figure*}
\begin{center}
\hbox{
\psfig{figure=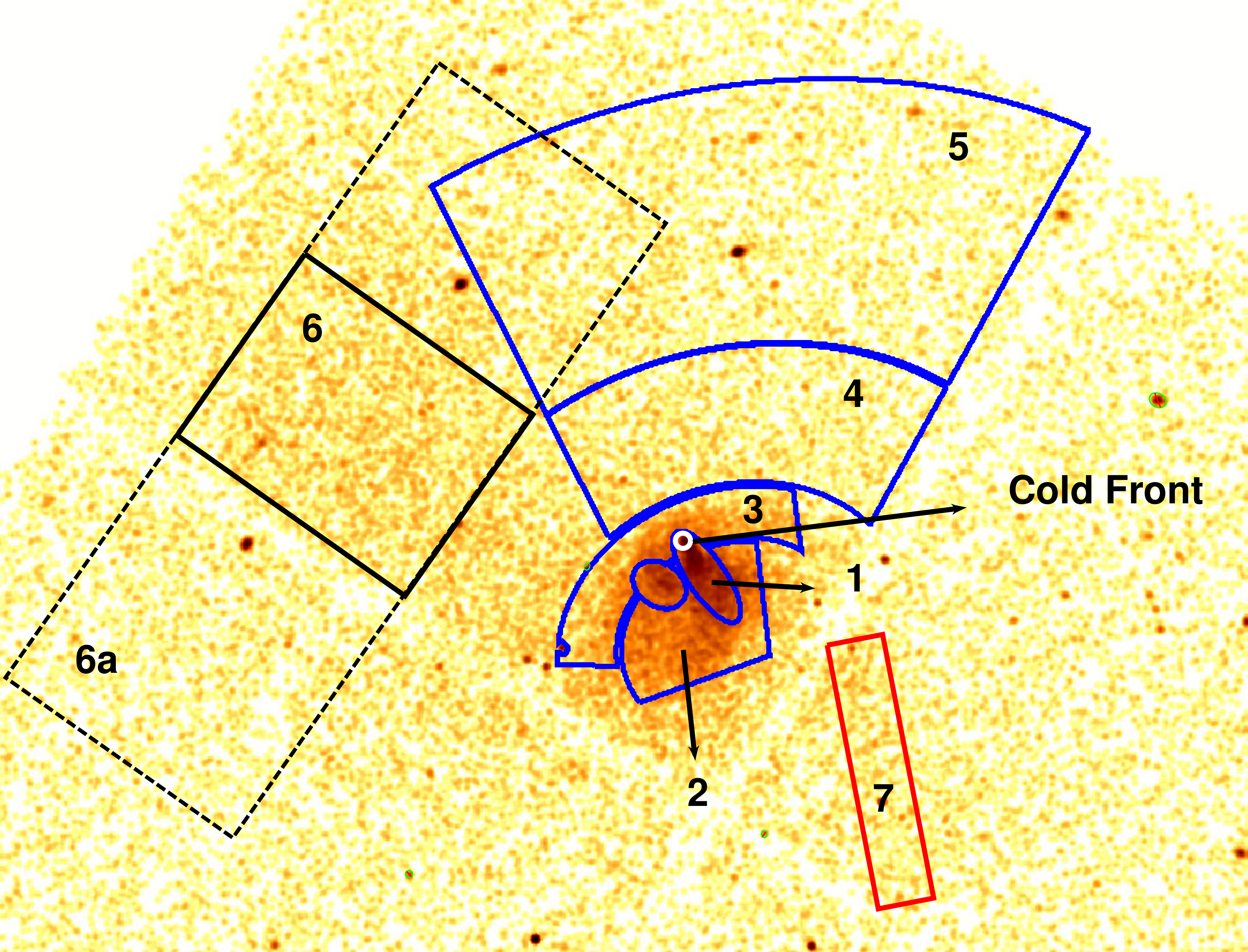,width=0.5\textwidth}
\psfig{figure=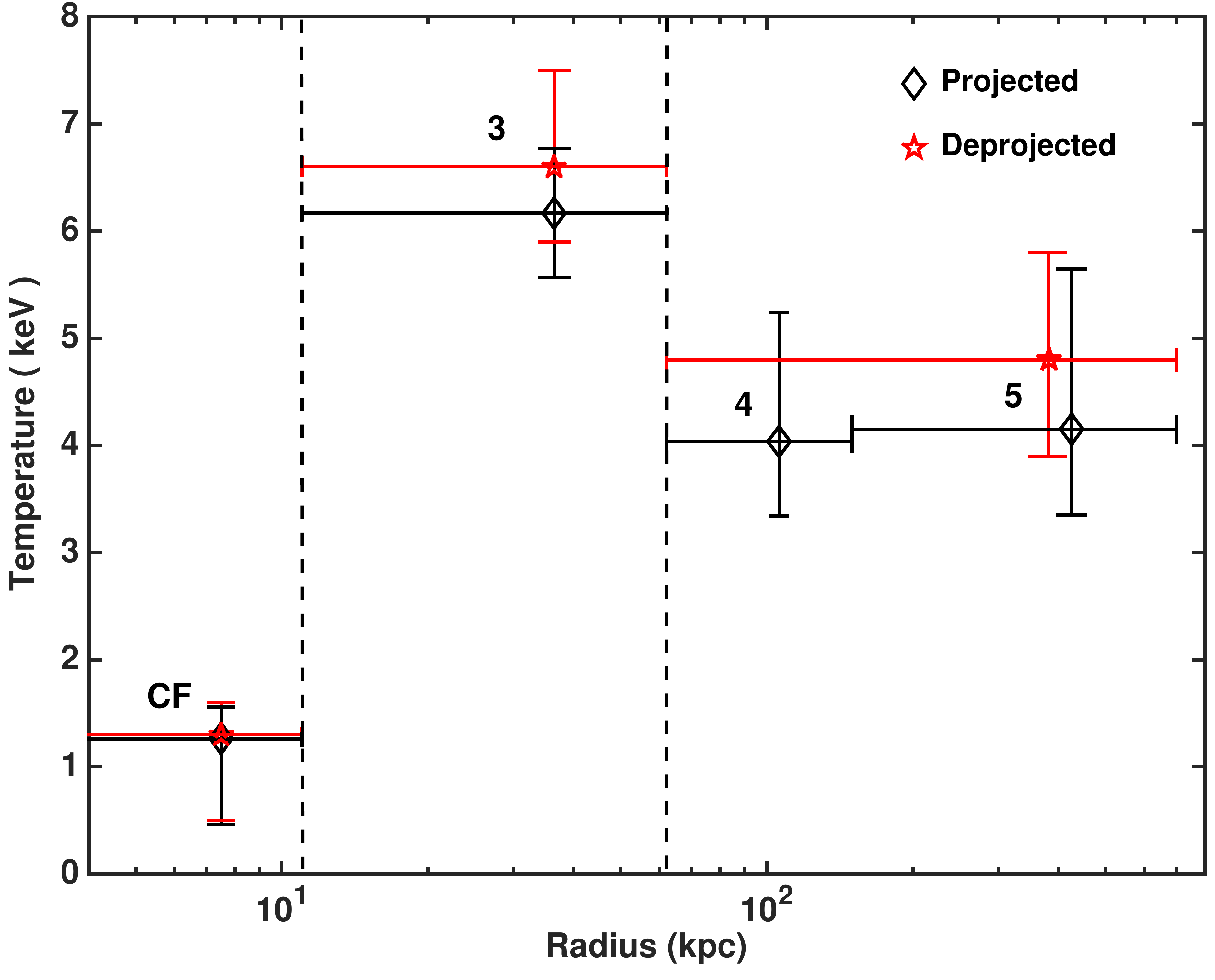,width=0.48\textwidth}
}
\caption{Left: Regions used for spectral analysis. Regions 4 and 5 are selected to avoid the EE. The ellipse around BCG 1 was excluded from the analysis. Region 6a covers both sides of the region 6 (the dashed line). Right: The projected (black) and deprojected (red) temperatures in regions CF, 3, 4 and 5, across the cold front and the shock front (see section 4 for the detail of deprojection).}
\end{center}
\end{figure*}

The key merger features identified in section 3.1 are the main subjects of our spectral studies. The regions for spectral analysis are shown in Fig. 5. The following factors are considered to define regions: 1) Each region should have at least 500 spectral counts; 2) the regions are shaped to follow surface brightness edges, covering their visible angular extent. Region 1 contains the X-ray tail of 3C89. Region 2 contains the ICM of the main cluster. The outer edge is enclosed by region 3. Regions 4 and 5 cover the region beyond the outer edge. We excluded the EE and the main cluster core. The spectrum from each region was extracted and grouped to give a minimum of 20 counts per channel (more for faint regions). A corresponding background spectrum was extracted from the normalized stowed data. Weighted response files were generated. The CXB model obtained by fitting the local background (section 2) was scaled to the area of each region and subtracted. XSPEC was used to fit an absorbed thermal model \citep[APEC;][]{s6} to spectra from both observations simultaneously. For the regions with fewer counts, the abundances were fixed close to the best-fit value obtained in nearby brighter regions. The temperature and normalization were left free. The best-fit parameters were obtained by minimizing the C-statistics. C-statistic does not provide a goodness-of-fit measure but the XSPEC version of the C-statistic is defined to approach $\chi^2$ in the case of large number of counts. The best-fit temperatures and abundance are reported with $1\sigma$ confidence intervals in Table 4. The reduced $\chi^2$ for spectral fits ranges from 0.9 - 1.2.

The projected (black diamond) and deprojected (red star) temperature values are plotted in Fig. 5. We calculated the deprojected temperature values assuming spherical symmetry, which is a reasonable assumption for wedge shaped regions across the shock front. The gas temperature beyond region 5, 3.1 keV, was derived from the azimuthally averaged temperature profile of the cluster. We assumed that the cluster have concentric spherical shells, each characterized by uniform temperature. Monte Carlo simulations were performed to populate the 3D shells. We then make an image where each pixel represents the number of events that fell within the 3D shell. The normalizing values for deprojection were estimated by applying respective masks to the simulated images. This method of geometrical deprojection allowed us to account for point sources, chip gaps and wedge shaped regions. Our estimated values for full annuli match with the volume - cylinder intersection fractions obtained using the ``onion-peeling" method. For the deprojection analysis, we combined regions 4 \& 5. The best-fit temperatures and the density model derived in the last section were multiplied to produce the electron pressure profile shown in Fig. 4 (the middle right panel).

Region 1 contains the X-ray brightest gas indicating the presence of a dense gas body with a sharp boundary. The spectrum of this region reveals the cold gas at a temperature of 3.8 $\pm$ 0.2 keV and an abundance  of $\sim$ 0.8 Z$_{\bigodot}$, before deprojection. This implies the presence of dense and metal-rich subcluster cool core. The head of the region 1 was isolated for the analysis of the cold front, as discussed later. Across the cold front, the drop in the surface brightness is accompanied by an increase in temperature. The higher abundance of the region 1 than most of the other regions implies gas of a different origin. Regions 2 and 3 contain the hottest gas of the cluster. The abundance in region 3 is poorly constrained. We fixed the abundance to 0.3 Z$_{\bigodot}$ which is the best-fit abundance in regions 4 and 2. The best-fit temperature changes by $<$ 3\% for $Z$ = 0.2 Z$_{\bigodot}$ - 0.4 Z$_{\bigodot}$. The temperature drops across the outer edge (S2) from 6.2 $\pm$ 0.6 keV to 4.1$_{-0.7}^{+1.2}$ keV in region 4. The gas temperature in the upstream side of the edge remains constant ($\sim$ 4 keV) for about 250 kpc from the edge. This implies that the contact discontinuity ahead of the cold front is a bow shock produced by the merger. 

When a body moves faster than the local sound speed, a bow shock is formed at some distance upstream. The surface brightness discontinuity at the bow shock implies a density jump of $\sim$ 2.2, which can also be estimated by the temperature jump across the shock front, $\sim$ 1.5 here. The Rankine-Hugoniot jump conditions, which relate the gas density and the temperature jump were used to derive an equivalent density jump. According to which, the temperature jump, $t  \equiv $ $T_{2}$ / $T_{1}$ is related to density jump $d$ $ \equiv $ $\rho$$_{2}$/$\rho$$_{1}$ by following equation,

\begin{equation}
   t=\frac{\xi - d^{-1}}{\xi - d},
\end{equation}
 
\noindent or, conversely, 

\begin{equation}
d^{-1}=\bigg[\frac{1}{4} \xi^{2} (t-1)^{2} + t\bigg]^{1/2}-\frac{1}{2} \xi (t-1),
\end{equation}

\noindent where we denoted $\xi$ $\equiv$ ($\gamma$ + 1)/($\gamma$ -1) $= 4$; here $\gamma$ = 5/3 is the adiabatic index for monoatomic gas. (\cite{m2}) and use indices 1 and 2 to denote quantity before and after shock.

\begin{table}
\protect\caption{Spectral Fits for Regions in Fig. 5. }
\begin{tabular}{|c|c|c|c|cl}
\hline 
Region & Spectral$^a$ & Temperature & Abundance$^b$\\& Counts & (keV)& Z$_{\bigodot}$ &\tabularnewline
\hline 
1 & 1201 & 3.8$_{-0.2}^{+0.2}$ & 0.8$_{-0.2}^{+ 0.2}$ \tabularnewline
\hline 
2 & 1750 & 6.0$_{-0.4}^{+0.4}$ & 0.3 $_{-0.1}^{+ 0.1}$ \tabularnewline
\hline 
3 & 975 & 6.2$_{-0.6}^{+0.6}$  & [0.3]  \tabularnewline
\hline 
4 & 824 & 4.1$_{-0.7}^{+1.2}$& 0.3$_{-0.3}^{+ 0.4}$ \tabularnewline
\hline 
5 & 1737 & 4.2$_{-0.8}^{+1.5}$ & [0.3]   \tabularnewline
\hline 
6 & 1191 & 6.3$_{-0.9}^{+1.0}$ & 0.3$_{-0.2}^{+ 0.2}$   \tabularnewline
\hline 
6a & 1853 & 3.4$_{-0.4}^{+0.6}$ & 0.3$_{-0.1}^{+ 0.2}$   \tabularnewline
\hline 
7 & 333 & 5.0$_{-1.8}^{+4.5}$ & [0.3]   \tabularnewline
\hline 
Cold Front (CF)$^{*}$ & 107 & 1.3$_{-0.8}^{+0.3}$ & [0.8]   \tabularnewline
\hline
\end{tabular}
$^a$ Spectral counts in the 0.7 - 7 keV energy band.\\
$^b$ The abundance in region 3, 5 and 7 was fixed to 0.3 Z$_{\bigodot}$ (as indicated by the square brackets), which is close to the best-fit abundance obtained in region 4 and 2. We notice a small change ($<$ 3\%) in best-fit temperature by changing $Z$ to 0.2 - 0.4 Z$_{\bigodot}$. We also fixed abundance in the cold front to 0.8 Z$_{\bigodot}$. The best-fit temperature changes by 0.02 keV for $Z$ = 0.5 - 1.0 Z$_{\bigodot}$. \\ 
$^{*}$ A circular region with a 12 kpc radius centered at the 3C89 nucleus. The CF region is at the front of the region 1 and does not overlap with the region 3 (Fig. 5).
\end{table}

With $T_{2}$ = 6.2 $\pm$ 0.6 keV and $T_{1} = $ 4.1$_{-0.7}^{+1.2}$ keV, the density jump $d$ = 1.7$_{-0.4}^{+0.6}$.  The Mach number of the shock is, $M = \upsilon$/c$_{s}$, where $v$ is the velocity of gas with respect to the shock surface and c$_{s}$ is the speed of sound in the pre-shock gas. The Mach number for the bow shock was found to be $M = 1.6^{+0.5}_{-0.3}$. Although the uncertainty in temperature is large, the Mach number derived from the temperature jump is consistent with the Mach number obtained from density jump. The sound speed in the pre-shock gas is, c$_{s}$ = 1.0$_{-0.1}^{+0.2}$ $\times$ 10$^3$ km sec$^{-1}$. Therefore, the shock velocity is, $v$ = 1.6$_{-0.3}^{+0.6}$ $\times$ 10$^3$ km sec$^{-1}$.

The small separation of the nucleus and the cold front, combined with the likely X-ray nuclear source, makes the spectral analysis inside the cold front challenging. The cold front properties were derived from a circle of 12 kpc radius centered at 3C89's nucleus.The spectrum from this region was fitted with the model TBabs(APEC + zTBabs $\times$ powerlaw) at fixed redshift. The powerlaw index was fixed at 1.7 and the abundance was fixed to the best-fit value obtained for region 1, 0.8 Z$_{\bigodot}$. The nH parameter of the zTBabs component couldn't be constrained due to low statistics. The best-fit temperature of $T_{\rm in}$ = 1.3$_{-0.8}^{+0.3}$ keV was used as the cold front temperature. $T_{\rm in}$ changes by $< 2\%$ for powerlaw index = 1.4 - 2.0 and $Z$ = 0.5 - 1.5 Z$_{\bigodot}$. We multiply the best-fit density model by the corresponding temperature to obtain the pressure distribution (see Fig. 4). 

The gas parameters at the stagnation point cannot be measured directly due to the small region size and the limited statistics. However, the pressure should be continuous across the cold front, so the cold front pressure gives the stagnation pressure. The ratio of the pressure in the free stream to the stagnation point is a function of the cloud speed \citep{l1} is given by

\begin{equation}
   \frac{p_{in}}{p_{out}}=\big(1 + \frac{\gamma -1} {2} M^2 \big)^{\frac{\gamma}{\gamma -1}} , M \le 1\\
\end{equation}

\begin{equation}
\frac{p_{in}}{p_{out}}= \big(\frac{\gamma + 1}{2} \big)^{\frac{\gamma + 1} {\gamma - 1}} M^2  \big[\gamma - \frac{\gamma -1} {2M^2} \big]^{\frac{-1} {\gamma - 1}} , M > 1
\end{equation}

\noindent where $M = v / c$ is the Mach number in the free stream and c is the sound speed in free stream. The subsonic equation (4) is derived from Bernoulli's equation while, the supersonic equation (5) also includes the
pressure jump at the bow shock.

The density just inside the cold front, calculated from the best-fit beta model, $n_{e,in}$ = 1.3 $\times$ 10$^{-2}$ cm$^{-3}$ was multiplied by temperature to obtain the pressure $p_{in}$ = 1.6 $\times$ 10$^{-2}$ keV cm$^{-3}$. We extrapolated density model outside the compression region (region III) to estimate the density just outside the cold front and multiplied it with outside temperature $T_{out}$ = 4.1 keV to obtain free stream gas pressure of $p_{out}$ = 0.37 $\times$ 10$^{-2}$ keV cm$^{-3}$. The pressure ratio between just inside and outside the cold front $p_{in}/p_{out}$ $\sim$ 4 corresponds to the Mach number $\sim$ 1.5. This is consistent with the Mach number of the shock front derived from the density and temperature jumps across the bow shock.

\begin{figure*}
 \centering
 \vspace{+2px}
 \hbox{\hspace{-10ex}\includegraphics[scale=.48]{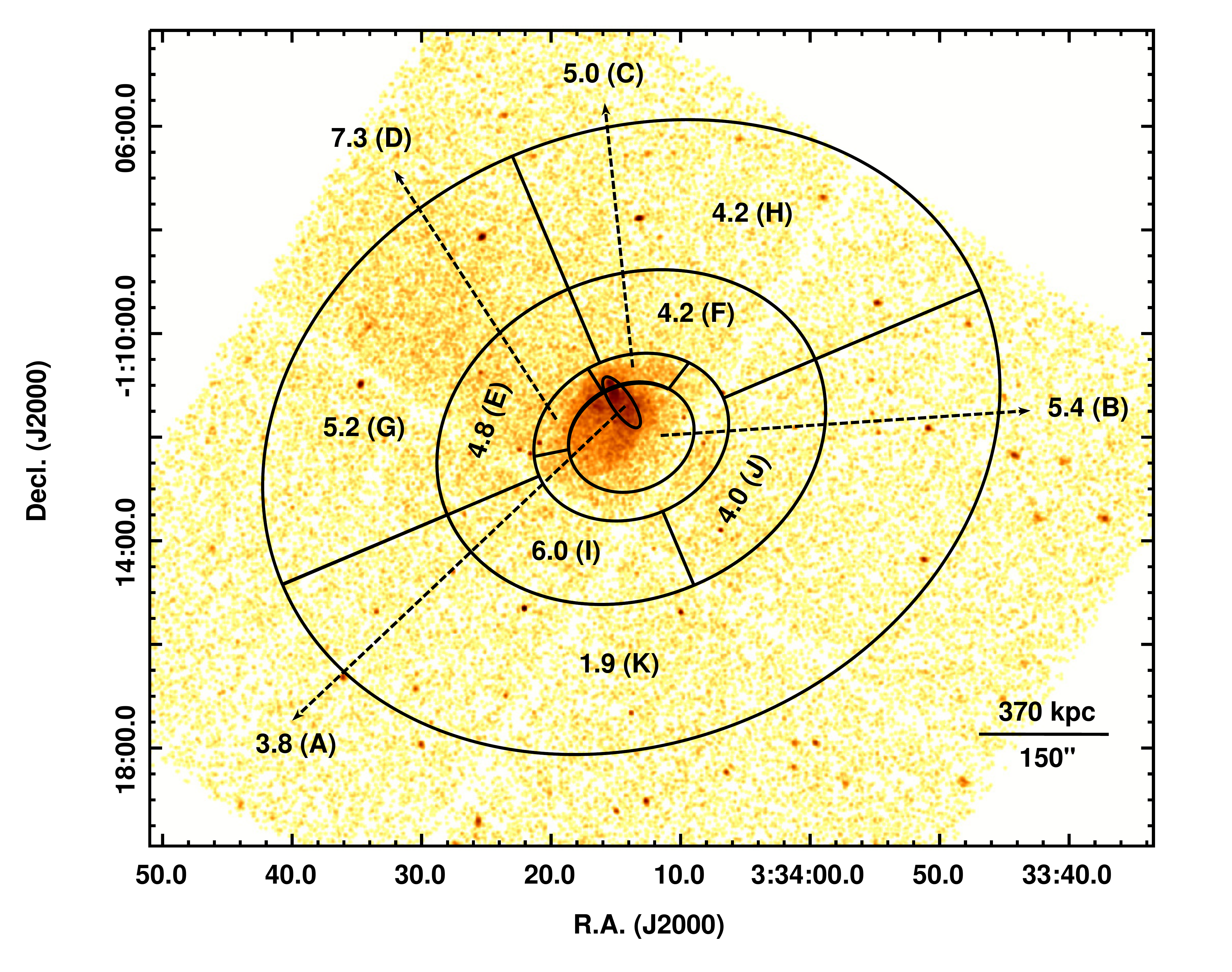}}
 \caption{Temperature distribution of the cluster (see Table 5 for detail). The numerical values are $kT$ in keV. The region to the opposite to C and D, has not been included in the spectral analysis.}
\end{figure*}

A projected temperature map was also generated in regions A to K (Fig. 6). The bins were made to closely follow the X-ray surface brightness edges. The abundance in regions with low counts was fixed at 0.3 Z$_{\bigodot}$. Table 5 shows the spectral properties of each region. The shock region (region 3 in Fig. 5) was divided into two segments C \& D. We note that the abundances in regions F and B, surrounding the shocked region are consistent with $\sim$ 0.3 Z$_{\bigodot}$. Thus, a fixed abundance of 0.3 Z$_{\bigodot}$ in regions C and D was used.

\begin{table}

\caption{Spectral Properties of Regions Shown in Fig. 6.}
\begin{tabular}{|c|c|c|c|cl}
\hline 
Region & Spectral$^a$ & Temperature & Abundance$^b$\\&Counts& (keV)& Z$_{\bigodot}$ &\tabularnewline\hline 
A &  1022 &3.8$_{-0.2}^{+0.2}$ & 0.8$_{-0.2}^{+0.2}$  \tabularnewline
\hline 
B &  1951 &5.4$_{-0.3}^{+0.4}$ & 0.3$_{-0.1}^{+0.1}$  \tabularnewline
\hline 
C &  362 & 5.0$_{-0.6}^{+0.7}$ & [0.3]  \tabularnewline
\hline 
D &  452 & 7.3$_{-1.0}^{+1.6}$ & [0.3] \tabularnewline
\hline 
E &  993 & 4.8$_{-0.5}^{+0.7}$ & 0.5$_{-0.2}^{+0.3}$  \tabularnewline
\hline 
F &  1678 & 4.2$_{-1.3}^{+2.6}$ & 0.3$_{-0.2}^{+0.3}$  \tabularnewline
\hline 
G &  2659 & 5.2$_{-0.6}^{+0.6}$ & 0.5$_{-0.2}^{+0.2}$  \tabularnewline
\hline 
H &  1678 & 4.2$_{-1.5}^{+2.3}$ & [0.3]  \tabularnewline
\hline 
I &  1162 & 6.0$_{-1.5}^{+2.0}$ & [0.3]  \tabularnewline
\hline 
J &  853 & 4.0$_{-0.9}^{+1.5}$ & [0.3]  \tabularnewline
\hline 
K &  3279 & 1.9$_{-0.3}^{+0.6}$ & [0.3]  \tabularnewline
\hline 
\end{tabular}
\\
$^a$ Spectral counts in the 0.7 - 7 keV energy band.\\
$^b$ The abundance in regions with limited statistics was fixed to 0.3  Z$_{\bigodot}$, as indicated by the square brackets.
\end{table}

\section{The 3C89 AGN}

3C89 is a classical wide angle tail radio galaxy. Its radio luminosity at 408 MHz, $L_{\rm 408 MHz}$, is of the order of 4$\times$10$^{26}$ W/Hz, among the highest for FR-I radio galaxies\citep[][]{z2}.  Its 1.4 GHz luminosity, 1.6 $\times$ 10$^{26}$ W Hz$^{-1}$, is comparable to that of many FR-II radio galaxies.  However, only weak emission lines are visible in its optical spectrum, that appears as that of a normal elliptical galaxy \citep[][]{b4}.  This was also confirmed by follow-up spectroscopic observation that allowed us to determine the value of the log [O III]/H$\beta$ = 0.03, placing 3C89 in the area of the optical spectroscopic diagrams populated by normal 3CR radio galaxies and well separated from the extremely low-excitation galaxies \citep[][]{c2}.  Thus, we can conclude that 3C89 is more similar to a normal FR-I radio galaxy than an FR-II.

The X-ray core is detected at 2.4$\sigma$ level in the 2 - 8 keV energy range. Assuming the typical photon index of 1.7, the 2 - 10 keV luminosity of the X-ray core is of the order of 6 $\times$ 10$^{41}$ erg sec$^{-1}$. The VLA archival data available trough the NRAO VLA Archive Survey (NVAS\footnote{http://archive.nrao.edu/nvas/}) database were examined. The 3C89 compact radio core is clearly detected in both 1.5 GHz and 8.4 GHz radio images, but no signatures of radio jet knots are visible close to the nuclear region. The curved extended structure (hereafter indicated as plumes or lobes) could be attributed to the impact of the high density ICM that confines the radio emitting
material arising from the radio jets. Two bright knots are visible in the 1.5 GHz radio image (e.g., Fig. 2) lying at distances of $\sim$ 50 kpc (SE direction) and at 40 kpc (W direction) from the core, but the presence of extended radio emission trailing them to the south (i.e. plumes) distinguishes them from typical FR-II hotspots.

Radio AGN can have a significant impact on the surroundings through jets. Bending of radio jets can be used to constrain some properties of the jets, e.g., the total kinetic power and the particle density \citep[e.g.,][]{o4}. From time-independent Euler's equation, we have: 

\begin{eqnarray}
\frac{P_{ram} R} {r_{j}} = w \Gamma^2 \beta^2;
\end{eqnarray}

\noindent where $w =  \rho_{j}  v_{j}^{2} + 4P_{min}$ is the relativistic enthalpy, $P_{min}$ is minimum synchrotron pressure, $\rho_j$ is the particle density in the jet, $v_j$ is the velocity of the jet flow, $P_{ram}$ is the ICM ram pressure, $r_{j}$ is the radius of the jet and $R$ is the radius of curvature of the bent jet. For a non-relativistic jet, $P_{min}$ is small and we do not include in our calculations. Thus, the total kinetic power of the jet $L_{k}$ $\approx (\pi/2) r^{2}_{j} \rho_{j} v^{3}_{j}$. The ram pressure can be estimated by, $P_{ram} = \rho v^{2}$, where $\rho$ and $v$ are the ICM gas density and velocity respectively. Using above relations, and adopting  $n_{e,ICM}$ = 1.5 $\times$ 10$^{-3}$ cm$^{-3}$, $v$ = 1.6 $\times$ 10$^3$ km sec$^{-1}$ , $r_{j}$ = 3 kpc, $R$ = 50 kpc (obtained from radio image) and $v_{j}$ = $ 0.3$ $c$ , we find $P_{ram}$ = 7.3 $\times$ 10$^{-11}$ dyne cm$^{-2}$ and the total kinetic power $L_{k}$ = 1.5 $\times$ 10$^{45}$ erg sec$^{-1}$. Often, the jet velocity is measured at the flaring point which is close to the nucleus. The average beaming speed for WATs was estimated to be (0.3 $\pm$ 0.1) $c$ by \cite{h1} under the assumption that the jets are intrinsically symmetrical and that the jet sidedness ratio is due to relativistic beaming effects. However, there is no direct evidence of relativistic speeds in such jets. Additionally, WAT jets disrupt at larger distance \citep[e.g.,][]{o5} which may reduce the jet speed. For a jet speed between 0.01 $c$ - 0.3 $c$, we estimate the total kinetic power of the jet $L_k$ = (0.05 - 1.5) $\times$ 10$^{45}$ erg sec$^{-1}$. \cite{b2} established an empirical relation between cavity power and the total radio power from a sample of 24 radio galaxies, although the scatter is quite substantial. For 3C89's radio luminosity, the cavity power would be $\sim$ 6.0 $\times$ 10$^{44}$ erg sec$^{-1}$ from the relations by \cite{b2}, consistent with the estimate from jet bending.

\section{Additional merger features}

\subsection{The Eastern Extension (EE)}

The EE is located $\sim$ 250 kpc east of the cluster center. This looks like a localized feature. A projected azimuthally averaged radial profile of the cluster (Fig. 3) shows a rise in surface brightness at the region between 250 to 700 kpc from the cluster center. This implies the presence of dense gas in the region. We compare surface brightness distribution from the E region shown in Fig. 4 (the upper left panel) with and without the EE. The projected surface brightness profile of the region is shown in Fig. 4 (the lower left panel). The data points in black (stars) and red (diamonds) show the surface brightness distribution with and without the EE respectively. The figure shows the brightness drops by a factor of 2 when this X-ray bright region is excluded. The profiles are consistent with each other beyond $\sim$ 1000 kpc, so the EE has an extent of $\sim$ 450 kpc. The density enhancement of the EE was estimated by deprojecting surface brightness profiles assuming a spherical geometry. We estimate the drop in the density by a factor of $\sim$ 1.5 from an average density of 4.5 $\times$ 10$^{-4}$ cm$^{-3}$ when EE was excluded. 

The spectral analysis (Table 4) shows that the EE (region 6) is hotter than its surroundings (region 6a). Since the EE is also denser than its surroundings, it is overpressurized. Its rest-frame 0.5 - 2 keV luminosity is 7.7 $\times$ 10$^{42}$ ergs sec$^{-1}$. The EE hosts BCG 3 (Fig. 1). This suggests that the EE is associated with another infalling subcluster. This third merging subcluster could explain the overpressurized gas in the region. However, we did not detect any upstream shock in current data. Deeper observations with a wider field of view would help to resolve the nature of the EE.
 
\subsection{The Tail}

The \textit{Chandra} image (Fig. 1 and 2) shows a faint ``tail'' southwest of the cluster center, behind 3C89's tail. The count image with a 2D elliptical beta model subtracted (Fig. 2d) shows an excess in the position of the faint tail. As a faint feature, it is difficult to determine its full extent using the current data. The tail is clearly visible over a distance of $\sim$ 500 kpc from 3C89 and is $\sim$ 30 kpc wide. However, the unsharp masked image suggests that it may extend up to a distance of $\sim$ 1 Mpc from 3C89. We also plot the surface brightness distribution across this feature (Fig. 4). Its spectral properties (region 7 in Fig. 5) were also studied. The regions on either side of the tail was used for local background. The best-fit temperature was found to be 5.0$_{-1.8}^{+4.5}$ keV, for a fixed abundance of 0.3 Z$_{\bigodot}$. The gas surrounding the tail has a temperature of 4.0$_{-0.9}^{+1.5}$ keV. The deprojection of the surface brightness suggests a density rise in the tail by a factor of $\sim$ 2 compared to the average surrounding gas density of 5.1 $\times$ 10$^{-4}$ cm$^{-3}$. Here, the line of sight extent of the tail was taken to be equal to its narrow width on the plane of the sky. Thus, the tail is likely to be an overpressurized feature. Its rest-frame 0.5 - 2.0 keV luminosity is (8.1 $\pm$ 2.0) $\times$ 10$^{41}$ ergs sec$^{-1}$.

\section{Discussion}

The simplest interpretation of the data is that RXJ0334.2-0111 is experiencing a two-body merger near the plane of sky. The system exhibits a complicated merger morphology with substructure seen on the both sides of the merger axis. The X-ray morphology is elongated in the SW-NE direction with two density discontinuities. The inner discontinuity is a cold front surrounded by the hot gas of the main cluster. Since the infalling cool core is moving faster than the local sound speed, a bow shock is seen $\sim$ 50 kpc upstream.

The cold front in this cluster has a distinctly narrow shape with the radius of curvature of $\sim$ 10 kpc. The radius of curvature of the shock front is $\sim$ 20 times greater than that of the cold front, the highest ratio known among
merging clusters. In case of the bullet cluster, this ratio is $\sim$ 4 while in Abell 2744 it is about 8. Such a small cold front has been observed in other mergers such as the southern subcluster in Abell 85 \cite[e.g.,][]{i2} and the NGC 4839 subcluster in Coma \cite[e.g.,][]{n2}, but no shocks have been reported there. Compared with the these two cases with similar X-ray morphology, the merger of the 3C89 subcluster is in a later stage. It is possible that at the end of mergers in these systems, only the high density core of the original cool core will survive, with properties similar to embedded coronae found in many clusters \cite[e.g.,][]{s07}.

Perhaps the most exciting feature of the merger is the shock ($M \sim$ 1.6 from the temperature jump vs. $\sim$ 1.9 from the density jump). This implies a shock velocity of $\sim$ 1.6 $\times$ 10$^3$ km sec$^{-1}$. The sharpness of this edge indicates that the shock motion is almost perpendicular to the line of sight. Assuming that the infalling cluster is moving close to the shock velocity \citep[although it can be lower, e.g.,][]{s7,m7}, we estimate the core passage occurred just 50 Myr ago. This is 5 - 6 times shorter than the time scales estimated for the Bullet cluster and Abell 2146 \citep{m4,r3}. 

\subsection{Interaction of X-ray and Radio Plasma}

The cooling time for the ICM gas at the center of many galaxy clusters is less than 10$^{10}$ years \cite[e.g.,][]{e1,p1}. Thus, the cool gas should trigger an inward flow of material known as ``cooling flow'' \cite[][]{f3}. In contrast, observations found much less cold gas than expected \cite[e.g.,][]{e2,mn07,f12}, suggesting a source of heat at the center of galaxy clusters. The widely accepted candidate is the energy injected by radio AGN. The \chandra\ data confirmed the idea by showing X-ray plasma being displaced by radio sources located at the center of the cluster \cite[e.g.,][]{m00,f4,mn07,f12}. Therefore, X-ray observations are critical for measuring the non-radiative output from the AGN jets.

\begin{figure}
\centering
\hbox{\hspace{-3px} 
\psfig{figure=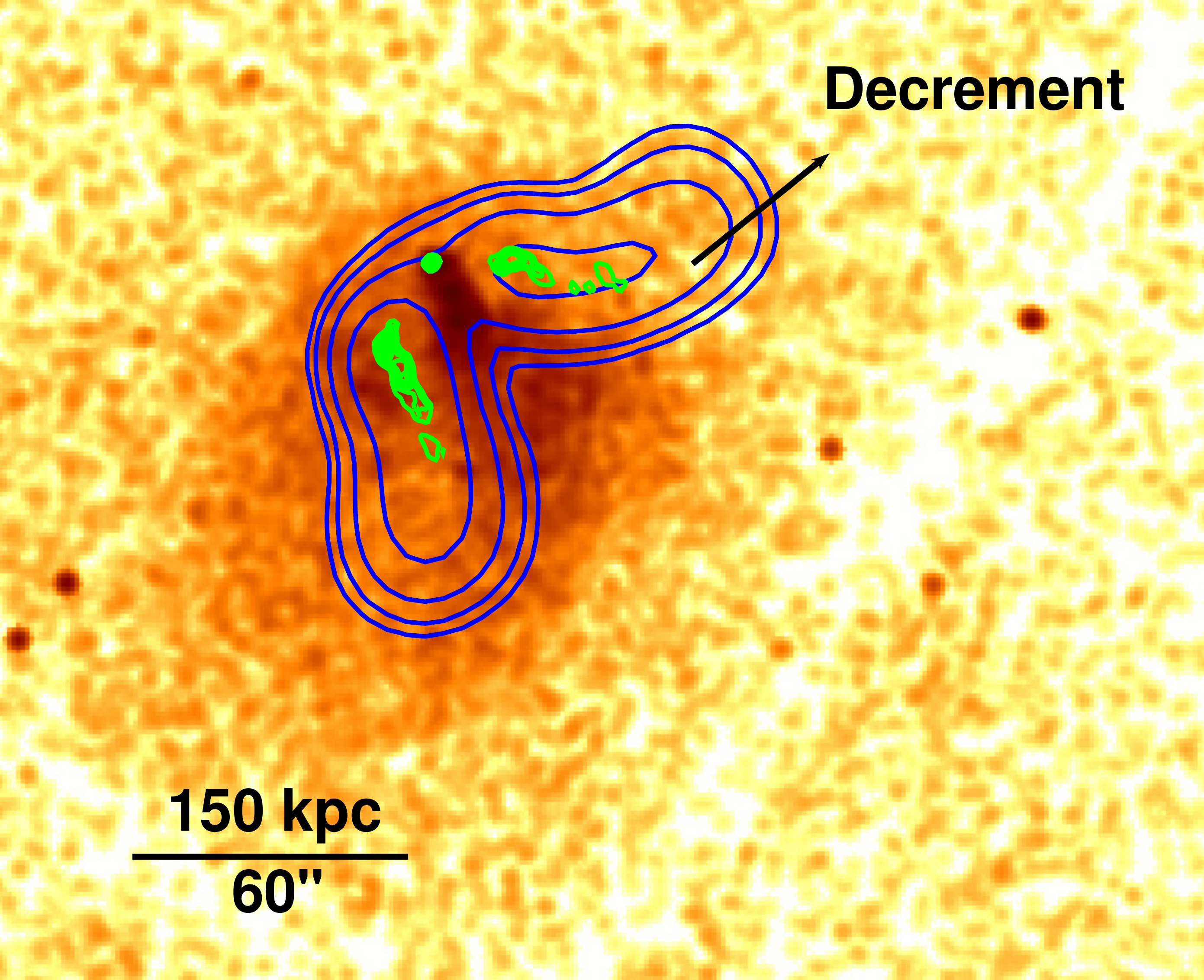,width=0.47\textwidth}
}
\caption{\chandra\ image of the cluster center overlaid with the 1.5 GHz \vla\ contours from two configurations, C in blue (a beam size of 14.8$''$ and a 0.124 mJy rms) and A in green (a beam size of 1.37$''$ and a 1.04 mJy rms). The radio lobe to the southeast is cutting through the X-ray decrement between 3C89 and BCG1, while the radio lobe to the west is bent upwards near the Decrement. We speculate that the interaction between X-ray and radio plasma is responsible for these features. }
\end{figure}

Fig. 7 shows the \chandra\ image of the cluster with 1.5 GHz \vla\ radio contours overlaid. Interestingly, the radio contours to the west are bent upwards near the Decrement. The Decrement is detected at 4.5$\sigma$ level in 0.7 - 2.0 keV energy range in the west. In the east, the radio jet is cutting through the main cluster core. However, the putative cavity is probably projected with the core of the main cluster, so the projected significance of the cavity is reduced. This suggests that the radio jets are working against the surrounding X-ray plasma to form a cavity. The total energy required to form a cavity is equal to its enthalpy, given by

\begin{equation}
   E_{cav}=\frac{\gamma}{\gamma -1}pV\\
\end{equation}

\noindent where $p$ is the pressure of the ICM gas surrounding the cavity, $V$ is the cavity volume, and $\gamma$ is the adiabatic index for the gas inside the cavity. For a relativistic gas, $\gamma$ = 4/3, and the total energy $E_{\rm cav}$ = 4$pV$ \cite[e.g.,][]{r8,mn07}. We assume the shape of the cavity is a cylinder of radius 30 kpc and length 100 kpc and determine, $E_{\rm cav}$ = 1.26 $\times$ 10$^{59}$ ergs. We assume that the bubble moves outwards from the AGN at the local sound speed and estimate the cavity's age, 9.7 $\times$ 10$^{7}$ years. The mean power required to create the cavity is, $P_{\rm cav}$ = 4.1 $\times$ 10$^{43}$ ergs sec$^{-1}$. A similar amount of power is expected at the other side of the radio AGN. The total cavity power estimated from the possible X-ray cavity is close to the jet power estimated from jet bending in section 5, especially if the jet speed is nonrelativistic. On the other hand, the possible X-ray cavity is small and may not contain the radio lobes, which will make our estimate of cavity power too low.

\subsection{Merger Scenario}

The cluster contains three luminous galaxies whose 2MASS $K_{\rm S}$ band luminosities are comparable to that of a typical BCG. The location of two galaxies (BCG 1 \& 2 in Fig. 1) are consistent with a merger scenario between two subclusters, where BCG 2 is the BCG of the infalling subcluster and BCG 1 is considered the cD galaxy of the main cluster. However, this merger scenario does not readily explain the EE, where BCG 3 is located. During the merger, gas stripped from the outskirts of the infalling subcluster may be left along an arcing trajectory, to form the tail-like feature. A cartoon shows the merger features in RXJ0334.2-01111 in Fig. 8.

\begin{figure}
\centering
\hbox{\hspace{-5px} 
\psfig{figure=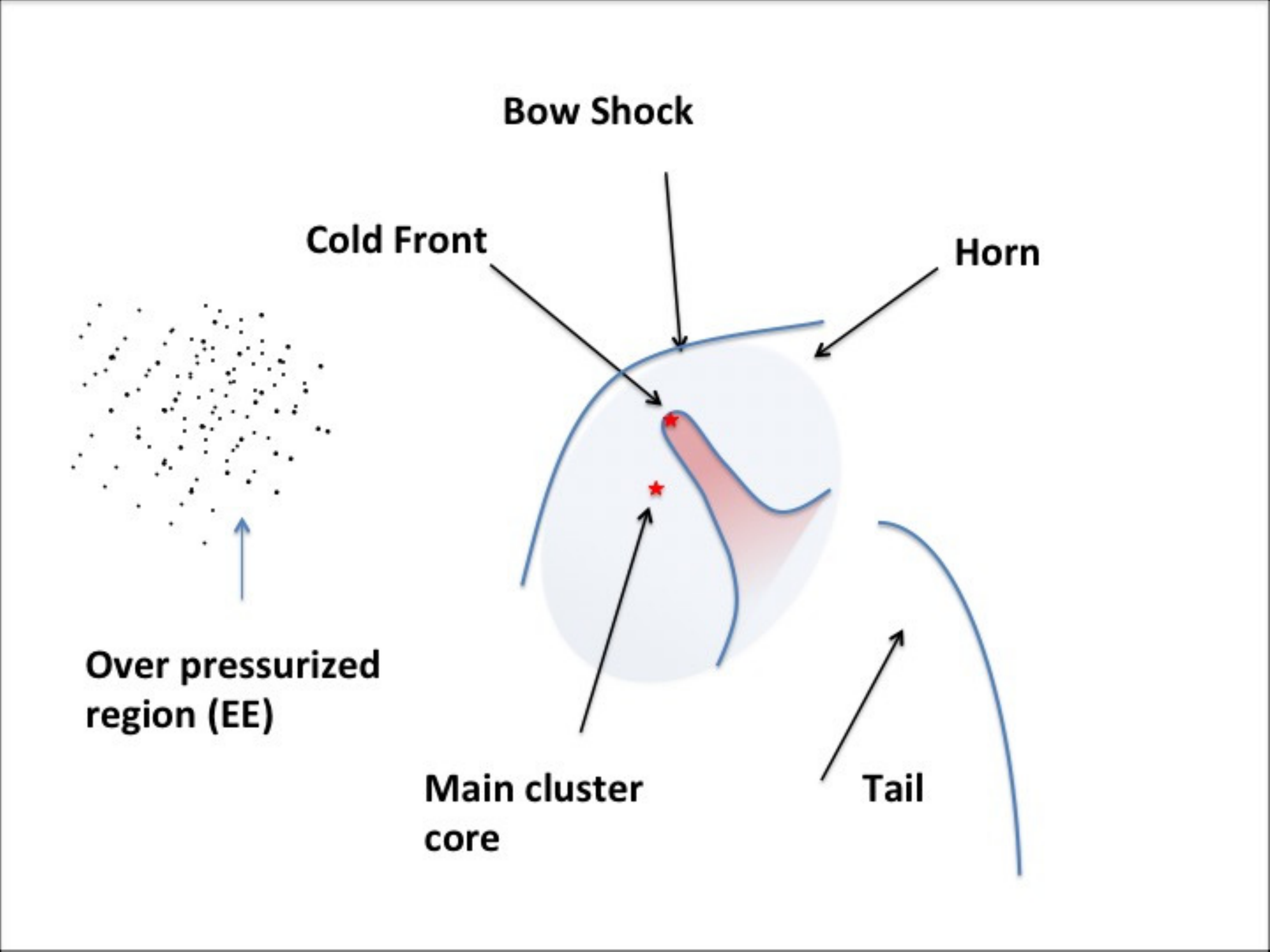,width=0.48\textwidth}
}
\caption{A cartoon showing the merging features in RXJ0334.2-0111. The image is not to scale. }
\end{figure}

A bow shock is formed when the infalling gas body moves faster than the local speed of sound. In some cases, a shock is also observed in the reverse direction e.g., Abell 2146 \citep{r3} and Abell 2219 \citep{r6}. The shocks in Abell 2146 have Mach number $\sim$ 2.3 (downstream) and $\sim$ 1.6 (upstream). In Abell 2219, merger has formed multiple edges which are consistent with the Mach number ranging from 1.1-1.2. The bow shock in RXJ0334.2-0111 is quite comparable to shocks found in other merging galaxy clusters. However, we do not observe any shock front in the reverse direction. 

The distance between the stagnation point and the closest point on the bow shock (the shock ``stand-off'' distance $d_{\rm s}$) can be estimated using the approximate method described by \cite{m8} \citep[see][for more details]{s8}. The shape of the cold front can be approximated by a cylinder with radius $\sim$ 5 kpc. Moeckel's method predicts a bow shock at the distance of 40 - 80 kpc from the cold front for Mach number $M$ = 1.6 - 2.0. This is in agreement with the observed distance of $\sim$ 50 kpc. On the other hand, for a rigid sphere, the ratio of the shock stand-ff distance, $d_{\rm s}$, to the radius of curvature of the cold front $R_{cf}$ depends on the Mach number $M$.  Fig. 9 (dashed line) shows the value of $d_{\rm s} / R_{cf}$ as function of $ (M^2 - 1)^{-1}$ \citep[][]{s9}. The data points show observed ratios, $d_{\rm s} / R_{cf}$, for known shocks located ahead of a cold front (Bullet Cluster \citep{m12}, Abell 754 \citep{m1}, Abell 520 \citep{m3}, Abell 2146 \citep{r3}, Abell 2744 \citep{o3}, RXJ0751.3+5012 \citep{r2}. We find that most of the clusters (except for Abell 754) don't agree with the model.  A possible explanation is that the infalling cool core constantly being stripped is not a rigid body.  Stripping continuously reduces the core radius and increases $d_{\rm s}$ at the same time.  We find that the $d_{\rm s} / R_{cf}$ ratios can be reduced to agree with the model, if the core radius is increased by a factor of 2 - 3 and $d_{\rm s}$ is decreased accordingly.

\begin{figure}
\centering
\hbox{\hspace{-15px} 
\psfig{figure=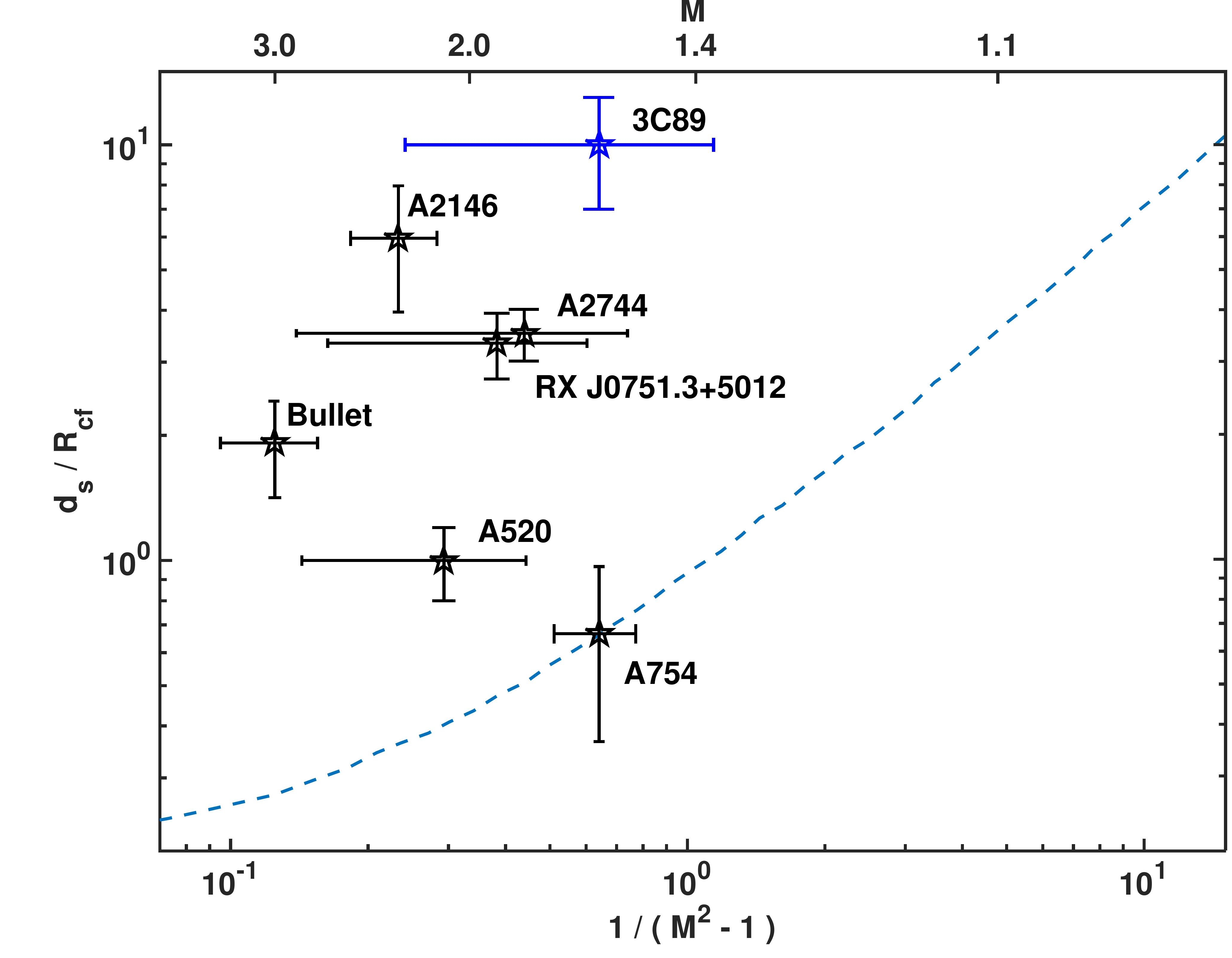,width=0.5\textwidth}
}
\caption{The ratio of the stand-off distance of the shock $d_s$ to the radius of curvature $R_{cf}$ of the stagnation region of the cold front, as a function of $ (M^2 - 1)^{-1}$ for all the known shock front/cold front combinations (3C89 included in blue) and for a rigid sphere from model (dashed line, see text for detail). The much larger ratios typically found in observations suggests that the core is continuously being stripped.}
\end{figure}

Many merging clusters hosting a robust bow shock exhibit diffuse radio emission e.g., Abell 520, Abell 754, Abell 2744 and the Bullet cluster \cite[e.g.,][]{m2}. The connection between merger substructures and diffuse radio emission has long been discussed \cite[e.g.,][]{m2,b1}. Interestingly, the diffuse radio sources are only found in clusters that show complex signs of mergers. For example, Abell 754 hosts a bow shock of similar strength (M $\sim$ 1.6) as RXJ0334.2-0111 and also exhibits diffuse radio emission in the outskirts \citep{m1}. This shock front coincides with the edge of non-thermal radio emission at 325 MHz \citep{m1}. Abell 2146 hosts both upstream and downstream shocks and surprisingly shows no evidence of extended radio emission \citep{r7}. 

For the clusters hosting bow shocks, we compare the shock strength with the corresponding system temperature of the cluster. In Fig. 10 for the Bullet Cluster \cite[$M \sim$ 3.0,][]{m12}, Abell 520 \cite[$M \sim$ 2.1,][]{m3}, Abell 2146 \cite[$M \sim$ 2.3 --downstream and $M \sim$ 1.6 --upstream,][]{r3}, Abell 521 \cite[$M \sim$ 2.4,][]{b5}, Abell 754 \cite[$M \sim$ 2.1,][]{m1}, Abell 2744 \cite[$M \sim$ 1.8,][]{o3}, RXJ0751.3+5012 \cite[$M \sim$ 1.9,]{r2}, Abell 2034 \cite[$M \sim$ 1.6,][]{o2}, Abell 2219 \cite[$M \sim$ 1.2 - 1.15,][]{r6}, Abell 1750 \cite[$M \sim$ 1.6,][]{b3}, Abell 3376 \cite[$M \sim$ 2.9,][]{a4} and Abell 3667 \cite[$M \sim$ 1.7,][]{f2}, the Coma Cluster \cite[$M \sim$ 1.8 --in the south and $M \sim$ 2.3 --in the west,][]{o6,a5,u1}. The Mach numbers stated above are consistent with the density and the temperature jump observed in each case. The significant of shock detection varies from case to case. In our study we don't include shock that are less significant e.g., the X-ray shock in the Sausage Cluster \cite[e.g.,][]{o7}. We divide the clusters into two categories with and without the diffuse radio emission (DRE). The line of equal shock velocities are marked with the dashed curves at $v$ = \{1, 2, 3, 4\} $\times$ 10$^3$ km sec$^{-1}$. Fig. 10 shows that the clusters without diffuse radio emission tend to have lower system temperature than clusters with diffuse radio emission.

\begin{figure}
\centering
\hbox{\hspace{-5px} 
\psfig{figure=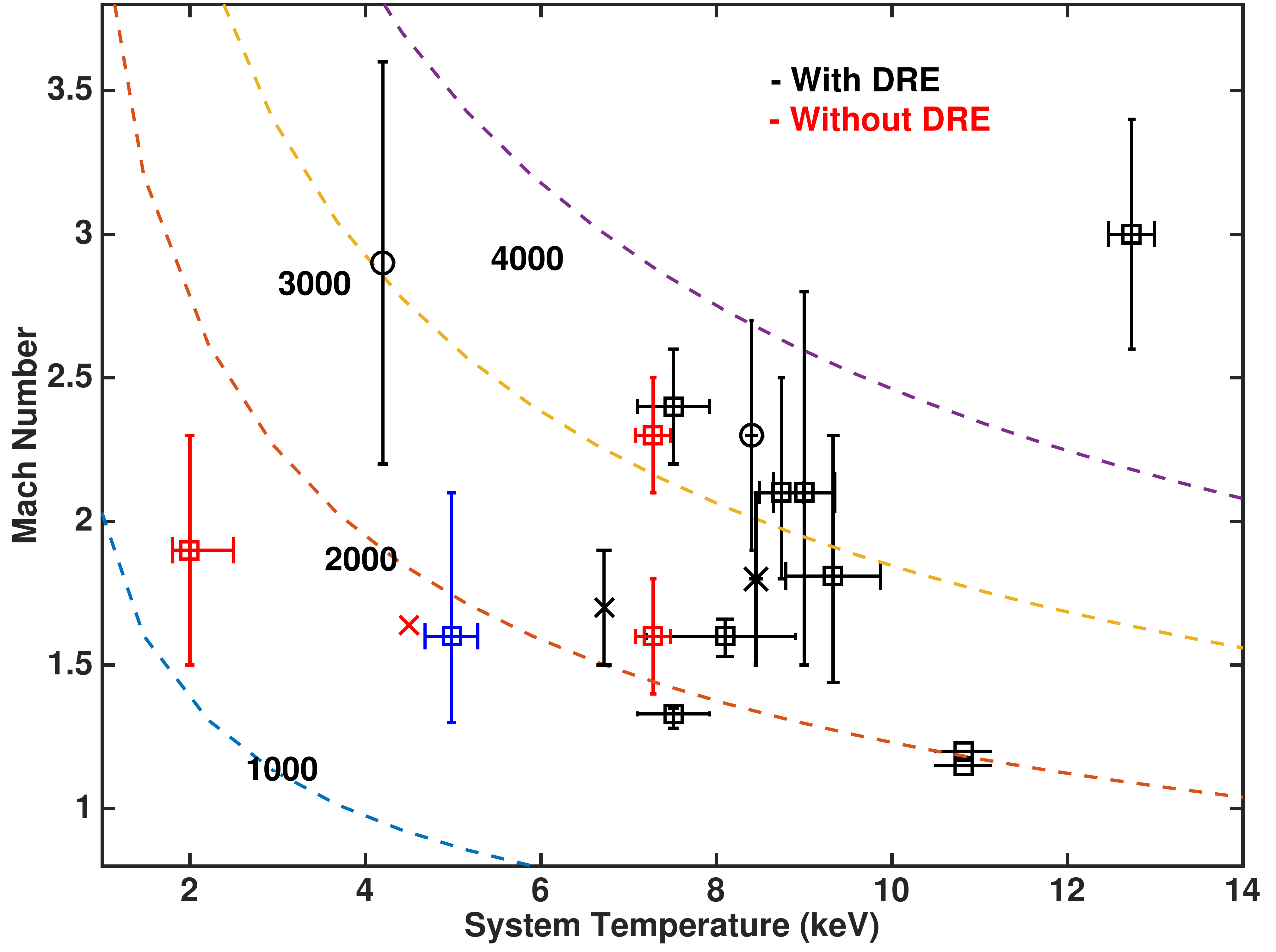,width=0.485\textwidth}
}
\caption{The Mach number vs. cluster temperature for known cluster shocks (see section 7.2 for detail). The shock discovered in this work is shown in blue. The clusters with and without diffuse radio emission (DRE) are shown in black and red respectively. The data points from \chandra, \xmm\ and {\em Suzaku} are displayed using square, ``X'', and ``O'' respectively.  The dashed lines are of equal shock velocities from 1000 to 4000 km sec$^{-1}$. The shock front discussed in this paper is highlighted by a blue data point. Note: not all works provided uncertainties for Mach numbers.}
\end{figure}

The lack of extended radio emission in RXJ0334.2-0111 could be due to several factors. Any detection of low surface-brightness radio features is hindered by the strong radio source 3C89 ($\sim$ 3 Jy at 1.4 GHz) and would require data with a high dynamic range. Detection of the diffuse radio emission is also easier at low radio frequencies than at 1.4 GHz. The other possibility is the relatively low mass of the cluster. In a study on the {\em GMRT} radio halo cluster sample, \cite{c1} identified the majority of clusters without a radio halo as having low X-ray luminosities and therefore low masses that are similar to RXJ0334.2-0111's.

\section{Conclusion}

The new 66 ksec \textit{Chandra} observation of RXJ0334.2-0111 revealed a merger between at least two subclusters. The image (Fig. 1) shows complex merging features. The primary result of this paper focuses on two surface brightness edges found in the X-ray observations. In addition to these edges, we found additional merger features such as the EE and a ``tail'' like feature. The following are the key conclusions of this paper.

1) The bright inner surface brightness edge is a cold front. It is formed when the cool, dense  core of the infalling subcluster passes through the hot ambient medium. A tail of ram pressure stripped gas can be seen following the cold front over a distance of $\sim$ 70 kpc. The observed stagnation pressure ratio is consistent with an infalling velocity of $\sim$ 1.5 $\times$ 10$^3$ km sec$^{-1}$. The cool core remnant of the main cluster can be seen SE of the cold front.
 
2) An additional surface brightness edge $\sim$ 50 kpc ahead of the cold front has been observed. This is a bow shock with a Mach number of 1.6$_{-0.3}^{+0.5}$ and is visible over $\sim$  500 kpc in length. The shock velocity of 1.6$_{-0.3}^{+0.6}$ $\times$ 10$^3$ km sec$^{-1}$ is consistent with the infalling cloud velocity. From the projected distance between two clusters of $\sim$ 90 kpc, we find the closest approach occurred $\sim$ 50 Myr ago.
 
3) The EE is an X-ray bright region located $\sim$ 250 kpc east of the cluster center. The region contains overpressurized gas at relatively high temperature ( $\sim$ 5.5 keV) and higher density than the surroundings. A massive galaxy (BCG 3 in Fig. 1) is likely the dominant galaxy of a third infall subcluster that is responsible for the EE.

4) A long, faint ``tail'' is observed to the west of 3C89, but the connection with 3C89's bright X-ray tail is not clear. It is clearly visible over a distance of $\sim$ 500 kpc and may extend up to $\sim$ 1 Mpc. The spectral analysis of this region suggests it is also an overpressurized region. This feature is faint in the current data so deeper observations are required to better understand it.

5) The radio lobes of the strong WAT source 3C89 are likely interacting with the X-ray gas, revealed by the X-ray decrement in the positions of radio lobes. We estimate the mechanical power required to create the Decrement-like cavity to be $\sim$ 10$^{44}$ ergs sec$^{-1}$, close to the the total jet power estimated from jet bending.

6) From the comparison between the ratio of the stand-off distance ($d_s$) to the radius of curvature of the cold front ($R_{cf}$) and the shock Mach number, it is found that most merger shocks in the clusters (apart from Abell 754) do not agree with the model results for a rigid sphere. We suggest that stripping continuously reduces the cool core radius and increases $d_s/R_{cf}$ ratio accordingly.  

The shock front in RXJ0334.2-0111 is only $\sim$ 50 kpc from the center of the cluster. Therefore, it has yet to propagate through the outskirts of the system. The bow shock is located ahead of the WAT radio galaxy, 3C89, the only WAT in a known merging galaxy clusters. Since the core passage occurred only $\sim$ 50 Myr ago, RXJ0334.2-0111 provides an excellent opportunity to study the effects of mergers on the cool core survival, jet bending etc. The substructures such as the Tail and the EE are unique to this system. Deeper X-ray observations and low frequency radio observations will help understand the significant questions raised by the current data.

\section{Acknowledgements}

We thank Maxim Markevitch and Marios Chatzikos for discussion and comments.
Support for this work was provided by the National Aeronautics and Space Administration (NASA) through \textit{Chandra} Award GO2-13160A and GO2-13102A issued by the \textit{Chandra} X-ray Observatory Center, which is operated by the Smithsonian Astrophysical Observatory for and on behalf of the National Aeronautics Space Administration under contract NAS8-03060.
This work was also supported in part by the NASA ADAP award NNX14AI29G. Basic research in radio astronomy at the Naval Research Laboratory is supported by 6.1 Base funding.
This research has made use of data and/or software provided by the High Energy Astrophysics Science Archive Research Center (HEASARC), which is a service of the Astrophysics Science Division at NASA/GSFC and the High Energy Astrophysics Division of the Smithsonian Astrophysical Observatory. 
The NRAO \vla\ Archive Survey image was produced as part of the NRAO VLA Archive Survey, (c) AUI/NRAO.


\begin{thebibliography}{99}


\bibitem[\protect\citeauthoryear{Akamatsu et al.}{2011}]{a4} Akamatsu H. et al, 2011,
PASJ, 64, 49
\bibitem[\protect\citeauthoryear{Akamatsu et al.}{2013}]{a5} Akamatsu H. et al, 2013,
PASJ, 65, 89
\bibitem[\protect\citeauthoryear{Anders \& Grevesse}{1989}]{a2} Anders E. \& Grevesse N., 1989,
Geochimica et Cosmochimica Acta 53, 197
\bibitem[\protect\citeauthoryear{Arnaud et al.}{1996}]{a1} Arnaud K., 1996, in Jacoby G., Barnes J., eds, ASP Conf. Ser. Vol. 101, Astronomical Data Analysis Software and Systems V. Astron. Soc. Pac., San Francisco, p. 17



\bibitem[\protect\citeauthoryear{Belsole et al.}{2004}]{b3} Belsole E. et al., 2004, A\&A, 
415, 821
\bibitem[\protect\citeauthoryear{B{\^i}rzan et al.}{2008}]{b2} B{\^i}rzan L. et al., 2008,
ApJ, 686, 859
\bibitem[\protect\citeauthoryear{Bonafede et al.}{2012}]{b1} Bonafede A. et al., 2012,
MNRAS, 426, 40
\bibitem[\protect\citeauthoryear{Bourdin et al.}{2013}]{b5} Bourdin H. et al., 2013,
ApJ, 764, 82
\bibitem[\protect\citeauthoryear{Buttiglione et al.}{2010}]{b4} Buttiglione S. et al., 2010,
A\&A, 509, A6



\bibitem[\protect\citeauthoryear{Capetti et al.}{2013}]{c2} Capetti A. et al., 2013,
A\&A, 551, A55

\bibitem[\protect\citeauthoryear{Cassano et al.}{2010b}]{c1} Cassano R. et al., 2010b,
ApJ, 721, L82
\bibitem[\protect\citeauthoryear{Canning et al.}{2015}]{r6} Canning R. et al., 2015,
arXiv:1505.05790



\bibitem[\protect\citeauthoryear{Edge et al.}{1992}]{e1} Edge A. et al., 1992,
MNRAS, 258, 177
\bibitem[\protect\citeauthoryear{Edge}{2001}]{e2} Edge A., 2001,
MNRAS, 328, 762



\bibitem[\protect\citeauthoryear{Fabian}{1994}]{f3} Fabian A. C., 1994,
ARA\&A, 32, 277
\bibitem[\protect\citeauthoryear{Fabian et al.}{2000}]{f4} Fabian A. C. et al., 2000, MNRAS, 318, L65
\bibitem[\protect\citeauthoryear{Fabian et al.}{2012}]{f12} Fabian A. C., 2012, ARA\&A, 50, 455 
\bibitem[\protect\citeauthoryear{Finoguenov et al.}{2010}]{f2} Finoguenov A., 2010,
ApJ, 715, 1143



\bibitem[\protect\citeauthoryear{Hardcastle \& Sakelliou}{2004}]{h1} Hardcastle M. \& Sakelliou I., 2004,
MNRAS, 349, 560


\bibitem[\protect\citeauthoryear{Ichinohe et al.}{2015}]{i2} Ichinohe Y. et al., 2015,
MNRAS, 448, 2971
\bibitem[\protect\citeauthoryear{Inogamov et al.}{1999}]{i1} Inogamov N. et al., 1999,
Astrophys. Space Phys. Rev., 10, 1



\bibitem[\protect\citeauthoryear{Jones \& Owen}{1979}]{o4} Jones T. \& Owen F., 1979,
ApJ, 234, 818



\bibitem[\protect\citeauthoryear{Kalberla et al.}{2005}]{k1} Kalberla P. et al., 2005,
A\&A, 440, 775  
\bibitem[\protect\citeauthoryear{Korngut et al.}{2011}]{k2} Korngut P. et al., 2011,
ApJ, 734, 10 
\bibitem[\protect\citeauthoryear{Kraft et al.}{2011}]{k3} Kraft R. et al., 2011, ApJ, 727, 41 




\bibitem[\protect\citeauthoryear{Landau \& Lifshitz}{1959}]{l1} Landau L., Lifshitz E., 1959, 
Fluid mechanics, Oxford, Pergamon Press



\bibitem[\protect\citeauthoryear{Macario et al.}{2011}]{m1} Macario G. et al., 2011,
ApJ, 728, 82 
\bibitem[\protect\citeauthoryear{Markevitch et al.} {1999}]{m5} Markevitch M. et al., 1999,
ApJ, 527, 545
\bibitem[\protect\citeauthoryear{Markevitch et al.}{2000}]{m9} Markevitch M. et al., 2000,
 ApJ, 541, 542
\bibitem[\protect\citeauthoryear{Markevitch et al.}{2002}]{m4} Markevitch M. et al., 2002,
ApJ, 567, L27
\bibitem[\protect\citeauthoryear{Markevitch}{2005}]{m12} Markevitch M., 2005,
Proceedings of the The X-ray Universe 2005 (ESA SP-604). 26-30 September 2005, El Escorial, Madrid, Spain. Editor: A. Wilson, p.723
\bibitem[\protect\citeauthoryear{Markevitch et al.}{2005}]{m3} Markevitch M. et al., 2005,
ApJ, 627, 733 
\bibitem[\protect\citeauthoryear{Markevitch \& Vikhlinin}{2007}]{m2} Markevitch M. \& Vikhlinin A., 2007,
Phys. Rept., 443, 1
\bibitem[\protect\citeauthoryear{Massaro et al.}{2012}]{m13} Massaro F. et al., 2012,
ApJS, 203, 31 
\bibitem[\protect\citeauthoryear{Massaro et al.}{2015}]{m14} Massaro F. et al., 2015,
ApJS, 220, 5 
\bibitem[\protect\citeauthoryear{McNamara et al.}{2000}]{m00} McNamara B., 2000, ApJ, 534, L135
\bibitem[\protect\citeauthoryear{McNamara \& Nulsen}{2007}]{mn07} McNamara B. \& Nulsen P. E. J., 2007, ARA\&A, 45, 117 
\bibitem[\protect\citeauthoryear{Million et al.} {2010}]{m6} Million E. et al., 2010
MNRAS, 407, 2046
\bibitem[\protect\citeauthoryear{Milosavljevic et al.} {2007}]{m7} Milosavljevic M., Koda J., Nagai D., Nakar E., \& Shapiro P., 2007
ApJ, 661, L131
\bibitem[\protect\citeauthoryear{Moeckel et al.} {1943}]{m8} Moeckel W. et al., 1943
NACA Technical Note 1921
\bibitem[\protect\citeauthoryear{Morandi et al.} {2015}]{m11} Morandi A. et al., 2015
MNRAS, 450, 2261



\bibitem[\protect\citeauthoryear{Neumann et al.} {2003}]{n2} Neumann D. et al., 2003,
A\&A, 400, 811
\bibitem[\protect\citeauthoryear{Nulsen et al.} {1982}]{n1} Nulsen P. et al., 1982,
MNRAS, 198, 1007



\bibitem[\protect\citeauthoryear{O'Dea \& Owen}{1985}]{o1} O'Dea C. \& Owen F., 1985,
AJ, 954, 972
\bibitem[\protect\citeauthoryear{O'Donoghue, Owen \& Eilek }{1990}]{o5} O'Donoghue A., Owen F., Eilek J., 1990,
ApJS, 72, 75
\bibitem[\protect\citeauthoryear{Ogrean et al.}{2013}]{o6} Ogrean G. et al., 2013,
MNRAS, 433, 1701
\bibitem[\protect\citeauthoryear{Ogrean et al.}{2014}]{o7} Ogrean G. et al., 2014,
MNRAS, 440, 3416
\bibitem[\protect\citeauthoryear{Owers et al.}{2011}]{o3} Owers M. et al., 2011,
ApJ, 728, 27
\bibitem[\protect\citeauthoryear{Owers et al.}{2014}]{o2} Owers M. et al., 2014,
ApJ, 780, 163


\bibitem[\protect\citeauthoryear{Peres et al.}{1998}]{p1} Peres C. et al., 1998,
MNRAS, 298, 416



\bibitem[\protect\citeauthoryear{Rafferty et al.}{2006}]{r8} Rafferty D. et al., 2006,
ApJ, 652, 216
\bibitem[\protect\citeauthoryear{Randall et al.}{2006}]{r5} Randall S. et al., 2005,
ApJ, 688, 208
\bibitem[\protect\citeauthoryear{Randall et al.}{2008}]{r1} Randall S. et al., 2008,
ApJ, 679, 1173
\bibitem[\protect\citeauthoryear{Roediger et al.}{2015}]{r4} Roediger E. et al., 2015,
ApJ, 806, 104
\bibitem[\protect\citeauthoryear{Russell et al.}{2011}]{r7} Russell H. et al., 2011,
MNRAS, 417, L1
\bibitem[\protect\citeauthoryear{Russell et al.}{2012}]{r3} Russell H. et al., 2012,
MNRAS, 406, 1721
\bibitem[\protect\citeauthoryear{Russell et al.}{2014}]{r2} Russell H. et al., 2014,
MNRAS, 444, 629



\bibitem[\protect\citeauthoryear{Schreier}{1982}]{s9} Schreier S., 1982,
Compressible Flow (New York: Wiley), 182-189

\bibitem[\protect\citeauthoryear{Shiparo}{1953}]{s8} Shiparo, A., 1953,
The Dynamics and Thermodynamics of Compressible Fluid Flow
\bibitem[\protect\citeauthoryear{Smith et al.}{2001}]{s6} Smith R., Brickhouse N., Liedahl D., \& Raymond J., 2001,
ApJ, 556, L91
\bibitem[\protect\citeauthoryear{Snowden et al.}{2008}]{s1} Snowden S. et al., 2008,
A\&A, 615, 658
\bibitem[\protect\citeauthoryear{Spergel \& Steinhardt}{2000}]{s2} Spergel D. \& Steinhardt P., 2000,
Phys. Rev. Lett., 84, 3760
\bibitem[\protect\citeauthoryear{Spinrad et al.}{1985}]{s3} Spinrad H. et al., 1985,
ASPP, 97, 932
\bibitem[\protect\citeauthoryear{Springel \& Farrar}{2007}]{s7} Springel V., \& Farrar G., 2007,
MNRAS, 380, 911
\bibitem[\protect\citeauthoryear{Sun et al.}{2002}]{s02} Sun M. et al., 2002, ApJ, 565, 867
\bibitem[\protect\citeauthoryear{Sun et al.}{2007}]{s07} Sun M. et al., 2007, ApJ, 657, 197
\bibitem[\protect\citeauthoryear{Sun et al.}{2009}]{s4} Sun M. et al., 2009, ApJ, 693, 1142


\bibitem[\protect\citeauthoryear{Uchida et al.}{2015}]{u1} Uchida Y. et al., 2015,
arXiv:1509.01901


\bibitem[\protect\citeauthoryear{Vikhlinin et al.}{2001}]{v1} Vikhlinin A., Markevitch M., \& Murray S., 2001, ApJ, 549, L47



\bibitem[\protect\citeauthoryear{Willingale et al.}{2013}]{w1} Willingale R. et al., 2013,
MNRAS, 431, 394


\bibitem[\protect\citeauthoryear{Zirbel \& Baum}{1995}]{z2} Zirbel E. \& Baum S., 1995, 
ApJ, 448, 521

\bibitem[\protect\citeauthoryear{ZuHone et al.}{2015}]{z1} ZuHone J. et al., 2015,
ApJ, 798, 90



\end{thebibliography}
\end{document}